\documentclass[twocolumn]{iopjournal} %iopjournal
\usepackage{fix-cm} % 新增此行，解决18pt字体缺失警告
\usepackage{microtype} % 优化英文断字，避免单词挤压难看
\usepackage{ragged2e} % 新增：提供\justifying命令
\AtBeginDocument{\justifying}% 文档加载完成后全局开启两端对齐
\usepackage{balance} % 均衡双栏底部，强烈建议加上
\usepackage{amsmath}

\begin{document}
\pagestyle{plain}
% \articletype{Paper} %	 e.g. Paper, Letter, Topical Review...
\twocolumn[
\title{Particle-resolved pathways to energetic-ion formation in a fluctuating low-current hollow-cathode plume}

\author{Baisheng Wang$^1$, Zilong Peng$^1$, Yinjian Zhao$^{1,*}$ and Zhongxi Ning$^1$}

\affil{$^1$  School of Energy Science and Engineering, Harbin Institute of Technology, Harbin, 150001, China}

\affil{$^*$Author to whom any correspondence should be addressed.}

\email{zhaoyinjian@hit.edu.cn}
\vspace{2mm}

\begin{abstract}
Energetic-ion formation in a low-current hollow-cathode plume is
investigated using experiments, self-consistent electrostatic
particle-in-cell (PIC) simulation, and particle-resolved analysis.
Retarding potential analyzer measurements show a substantial
energetic-ion population over discharge currents of $0.8$--$3.5$~A,
while probe measurements reveal broadband plume fluctuations.
Two-point phase-derived frequency--wavenumber measurements do not
resolve a continuous ion-acoustic dispersion branch within the
principal apparent-wavenumber interval. Because the inferred
wavenumber is obtained from a cross-spectral phase defined modulo
$2\pi$, the fluctuation diagnostics do not provide an unambiguous
modal attribution for the energetic-ion population.
A representative PIC plume, used as a qualitative kinetic reference,
likewise develops broadband time-dependent electrostatic fluctuations
together with a nonthermal energetic-ion population. Particle-resolved
analysis shows that the energetic outflow is dominated by ions generated
through ionization inside the plume, while source localization biases
access to distinct trajectory and escape families. Matched field
controls further show that time-averaged and frozen fields strongly
suppress access to high-energy trajectories relative to the full
time-dependent field over the analyzed interval.
At the single-particle level, ion kinetic-energy gain is determined by
electrostatic-field work accumulated along the actual trajectory, with
different escape families exhibiting distinct radial and axial work
contributions. These results establish a
source--trajectory--field--work pathway for energetic-ion formation
that can be identified without first assigning the fluctuating plume
to a unique resolved plasma mode.
\end{abstract}
\vspace{2mm}
\vspace{2mm}

(Some figures may appear in colour only in the online journal)
\vspace{2mm}
]

\section{Introduction}
\label{sec:introduction}

Hollow cathodes are widely used as electron sources in electric
propulsion systems, where they sustain the discharge and neutralize the
ion beam. Their operation involves strongly coupled electron emission,
ionization, plasma transport and particle--surface interactions,
producing highly nonequilibrium plasma conditions in both the cathode
and downstream plume
\cite{goebel2021plasma,lev2019recent}. Of particular concern is the
formation of energetic ions, which can enhance sputtering of the keeper,
cathode orifice, and surrounding surfaces and thereby affect cathode
lifetime
\cite{goebel2007potential,goebel2005energetic}.

Energetic ions well above the thermal ion-energy range have been
reported in hollow-cathode plumes using electrostatic energy analyzers,
retarding potential analyzers, and laser-induced fluorescence
measurements
\cite{williams2000laser,foster2005downstream}.
These observations have motivated several candidate mechanisms for
energetic-ion production. Stationary or slowly varying electrostatic
structures near the cathode orifice can accelerate ions through local
potential drops
\cite{patterson1999generation}, while collisional pathways involving
charge exchange and subsequent reionization have also been proposed
\cite{katz2006production}. A purely stationary acceleration picture,
however, has not accounted for all observations. In particular,
energetic ions have been observed together with pronounced
plasma-potential and density fluctuations even when a stationary dc
potential structure sufficiently large to explain the measured ion
energies was not identified
\cite{goebel2007potential}. These observations motivated increased
attention to time-dependent plume dynamics.

Hollow-cathode plumes support several classes of oscillations and
instabilities, including ionization-related oscillations arising from
the coupled evolution of the plasma, neutral population, and ionization
rate
\cite{georgin2020ionization,meng2019triggering,zhao2026review}.
Ion-acoustic turbulence (IAT) has received particular attention because
propagating ion-acoustic fluctuations have been resolved in high-current
LaB$_6$ hollow cathodes
\cite{jorns2014ion}. Subsequent studies have connected such fluctuations
with anomalous electron transport and the self-consistent electric-field
environment
\cite{jorns2016first,georgin2019correlation}, while energetic-ion
measurements and velocity-resolved diagnostics have provided evidence
linking ion-acoustic activity with ion energization under conditions
where a well-developed ion-acoustic response is present
\cite{jorns2014investigation,dodson2019measurements}.
The relation between energetic-ion production and ion-acoustic activity,
however, varies with operating condition and discharge regime.
Ion-energy distributions change substantially with discharge mode and
instability state
\cite{imai2022effect,wang2022presence,miao2025experimental}, and
energetic ions have also been reported in regimes dominated by
low-frequency discharge oscillations in which a well-developed
high-frequency ion-acoustic response was not clearly identified
\cite{wang2022presence}.

These observations illustrate a broader limitation of interpreting
energetic-ion formation from collective plasma measurements alone.
The coexistence of energetic ions and plasma fluctuations does not by
itself provide an unambiguous modal attribution for the energetic-ion
population. More importantly, measurements of final ion-energy
distributions and collective fluctuation spectra do not reveal where an
energetic ion was created, which trajectories were accessible from its
source region, how the evolving electric field modified its motion, or
where along the resulting trajectory energy was transferred to the
particle. The particle-level connection among source localization,
trajectory accessibility, temporal electric-field evolution, and
energy transfer therefore remains incompletely resolved.

Particle-in-cell (PIC) simulations provide a means of examining this
connection because particle and electric-field histories can be
resolved simultaneously. Previous kinetic simulations have demonstrated
the importance of ionization, nonequilibrium particle dynamics, and
transient electric-field structures in hollow-cathode plumes
\cite{luo2024coupled,zhao2025generation}. In particular, a previous
kinetic study using a two-dimensional electrostatic PIC model identified
large-amplitude, time-dependent potential structures associated with
charge separation and showed that the resulting self-consistent plume
could generate a substantial energetic-ion population
\cite{zhao2025generation}. That study addressed the collective evolution
of the kinetic plume, but did not resolve the particle histories
connecting ion birth, trajectory accessibility, temporal field exposure,
and energy transfer along individual trajectories.

Establishing these connections requires source-resolved particle
histories and controlled comparisons in which the prescribed particle
source or electric-field history can be varied independently. The
present work therefore addresses this complementary particle-level
question using a modified representative PIC configuration together with
source classification, controlled birth-position calculations, matched
field-history tests, and trajectory-resolved electric-field work.

In this work, energetic-ion formation in a low-current hollow-cathode
plume is investigated using experimental diagnostics, self-consistent
electrostatic PIC simulation, and particle-resolved analysis. The
experiment establishes the coexistence of energetic ions and broadband
plume fluctuations over discharge currents of $0.8$--$3.5$~A.
Two-point phase-derived frequency--wavenumber measurements are used to
characterize the accessible apparent-wavenumber interval and to assess
whether the observed fluctuations provide a clearly resolved modal
signature.
A modified representative PIC configuration, based on the kinetic
framework of Ref.~\cite{zhao2025generation}, then provides
self-consistent particle and electric-field histories for the present
analysis. The numerical case is constructed using characteristic scales
relevant to the experiment rather than as a point-by-point reconstruction
of a specific experimental operating condition. The particle-resolved
analysis is organized around a source--trajectory--field--work pathway:
particle histories identify the source composition of the energetic
outflow, controlled birth-position calculations quantify
source-dependent trajectory accessibility, matched field controls
isolate sensitivity to temporal electric-field evolution, and
trajectory-resolved work quantifies the radial and axial electric-field
contributions to particle energy gain.

\section{Methods}
\label{sec:methods}

\subsection{Experimental setup}
\label{sec:experimental_setup}

\subsubsection{Cathode and plasma diagnostics}
\label{sec:plasma_diagnostics}

\begin{figure*}[htbp]
    \centering
    \includegraphics[width=0.8\linewidth]
    {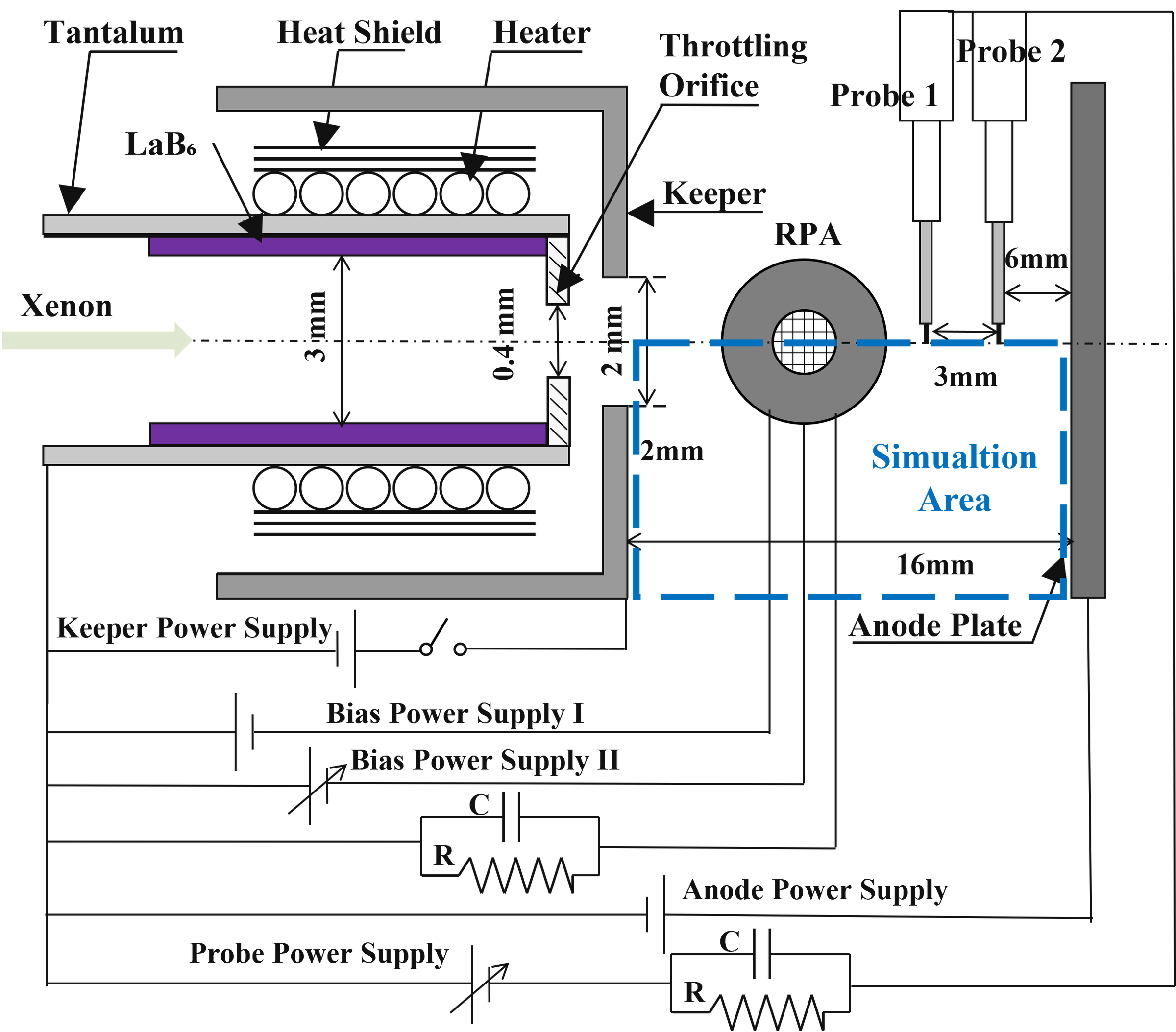}
    \caption{Schematic of the hollow-cathode experimental configuration,
    diagnostic arrangement, and corresponding PIC computational domain.
    The positions of the retarding potential analyzer (RPA) and the two
    synchronized Langmuir probes used for wave-propagation measurements
    are indicated. The blue dashed rectangle denotes the axisymmetric
    $r$--$z$ computational domain used in the PIC simulation.}
    \label{fig:cathode_experiment}
\end{figure*}

The experimental configuration, diagnostic arrangement, and the spatial
region represented by the PIC model are summarized in
Fig.~\ref{fig:cathode_experiment}. The experiment employed a
conventional orificed LaB$_6$ hollow cathode for Hall-thruster
applications. The 3~mm-diameter LaB$_6$ emitter was located inside a
tantalum tube upstream of a tungsten throttling plate. The cathode-orifice
and keeper-orifice diameters were 0.4~mm and 2~mm, respectively, and the
axial distance from the keeper face to the downstream anode plate was
16~mm.

After ignition, the keeper power supply was disconnected and the keeper
was allowed to float. Xenon was supplied at a fixed flow rate of
1.5~sccm, while the discharge current was varied from 0.8 to 3.5~A.
The gas pressure inside the tantalum tube was approximately
500--700~Pa, and the chamber background pressure was
$1.9\times10^{-2}$~Pa.

The plume was characterized using two Langmuir probes and a
retarding potential analyzer (RPA). The two synchronized probes used for
the wave-propagation measurements are indicated in
Fig.~\ref{fig:cathode_experiment}. Time-averaged plasma parameters were
obtained from swept-probe measurements, whereas negatively biased probes
were used to measure ion-saturation-current fluctuations. The background
electron temperature and ion density were measured on the discharge axis
6~mm downstream of the keeper face. The exposed probe tip was 2~mm long
and 0.5~mm in diameter.

The electron temperature was obtained from the exponential region of the
current--voltage characteristic of the swept single Langmuir probe,
following Ref.~\cite{chen2003mini}. The ion density was estimated from
the ion-saturation current using the thin-sheath approximation for a
cylindrical probe,

\begin{equation}
n_i =
\frac{i_{\mathrm{sat}}}
     {0.61 e A_p \sqrt{eT_e/m_i}},
\label{eq:ion_density_probe}
\end{equation}

where $i_{\mathrm{sat}}$ is the ion-saturation current, $A_p$ is the
probe collection area, $T_e$ is the electron temperature expressed in
electronvolts, $e$ is the elementary charge, and $m_i$ is the xenon-ion
mass.

For time-resolved measurements, the probes were biased at $-30$~V and
operated in the ion-saturation regime. 
Assuming that relative electron-temperature fluctuations are small
compared with relative density fluctuations, the ion-saturation-current
fluctuation provides a proxy for the relative density fluctuation
\cite{jorns2014ion},

\begin{equation}
\frac{\widetilde{i}_{\mathrm{sat}}}
     {\overline{i}_{\mathrm{sat}}}
\simeq
\frac{\widetilde{n}}{n_0}.
\label{eq:isat_density_fluctuation}
\end{equation}

Following the electrostatic-response approximation used in
Ref.~\cite{jorns2014ion}, the corresponding potential-fluctuation proxy
was estimated as

\begin{equation}
\widetilde{\phi}_{\mathrm{proxy}}
\simeq
T_e
\frac{\widetilde{i}_{\mathrm{sat}}}
     {\overline{i}_{\mathrm{sat}}},
\label{eq:potential_proxy}
\end{equation}

with $T_e$ expressed in electronvolts and
$\widetilde{\phi}_{\mathrm{proxy}}$ in volts.
Equation~\ref{eq:potential_proxy} is used only as a proxy for relative
potential fluctuations and does not constitute a direct measurement of
the instantaneous absolute plasma potential.

The RPA was positioned downstream of the keeper face and sampled ions
propagating in the radial direction, with its collection axis normal to
the cathode axis, as shown in Fig.~\ref{fig:cathode_experiment}. 
The analyzer employed a four-grid architecture following
Ref.~\cite{wang2022presence}. The first grid was electrically floating,
while the second and fourth grids were biased at $-24$~V to suppress
electrons. The third grid served as the retarding grid and was swept
from 0 to 250~V. The ion current reaching the collector was determined
from the voltage across a 500~k$\Omega$ collector resistor; a 10~nF
capacitor was connected in the collector circuit.

During each measurement, the retarding-grid voltage $V_r$ was swept
while the collector current $I_c$ was recorded. The differential
energy-per-charge distribution was obtained from

\begin{equation}
F(E/q_i)
\propto
-\frac{dI_c}{dV_r},
\label{eq:rpa_distribution}
\end{equation}

where $q_i$ is the ion charge. Accordingly, the RPA measures an
energy-per-charge distribution for the ions entering the analyzer
acceptance. Because the present PIC model contains singly charged
Xe$^+$ ions, the experimental retarding potential is reported below in
Xe$^+$-equivalent energy units, $E_{\mathrm{Xe^+}}=eV_r$. The RPA
itself does not resolve the ion charge state.

Before differentiation, the retarding-voltage and collector-signal
traces were smoothed, and a single increasing-voltage sweep was
interpolated onto a uniformly spaced $V_r$ grid. The collector-current
characteristic was then fitted with a cubic smoothing spline and
differentiated numerically using a finite-difference gradient. The
resulting derivative was further smoothed, negative values were set to
zero, and each RPA-derived distribution was normalized by its maximum.
The energetic-tail fraction reported below was defined as the integral
of the processed RPA distribution above a Xe$^+$-equivalent energy of
60~eV divided by its integral over the complete analyzed energy range;
the integrals were evaluated numerically using the trapezoidal rule.

\subsubsection{Wave propagation analysis}
\label{sec:fk_method}

Axial wave propagation was characterized using the two synchronized
Langmuir probes shown in Fig.~\ref{fig:cathode_experiment}, both
operated in the ion-saturation regime. The measured
ion-saturation-current fluctuations were used directly for the
cross-spectral frequency--wavenumber analysis following the
fixed-probe-pair method of Beall \emph{et al.} and its applications to
hollow-cathode plume measurements
\cite{beall1982estimation,jorns2014ion,miao2025experimental}.

Probe~1, located closer to the keeper, was used as the reference probe,
whereas Probe~2 provided the second fluctuation signal. For the phase
analysis, the signed probe separation was defined as
$
\Delta x
=
x_{\mathrm{Probe1}}-x_{\mathrm{Probe2}}
=
3.0~\mathrm{mm}
$.
For each time segment, the Fourier transforms of the two
ion-saturation-current fluctuation signals were denoted by
$X_{\mathrm{Probe2}}(f)$ and $X_{\mathrm{Probe1}}(f)$. The cross
spectrum and corresponding phase difference were defined as
$
P_{21}(f)
=
X_{\mathrm{Probe2}}(f)
X_{\mathrm{Probe1}}^{*}(f),
$ and
$\Delta\varphi(f)
=
\arg[P_{21}(f)]
$.
The corresponding phase-derived apparent axial wavenumber was defined as

\begin{equation}
k_{\mathrm{app}}(f)
=
\frac{\Delta\varphi(f)}{\Delta x}.
\label{eq:beall_wavenumber}
\end{equation}

Because $\Delta\varphi(f)=\arg[P_{21}(f)]$ is the principal
cross-spectral phase restricted to $(-\pi,\pi]$,
$k_{\mathrm{app}}$ is not, in general, a unique reconstruction of the
true axial wavenumber. For a single propagating component, the possible
true wavenumbers satisfy

\begin{equation}
k_{\mathrm{true}}(f)
=
k_{\mathrm{app}}(f)
+
\frac{2\pi n}{\Delta x},
\qquad
n=0,\pm1,\pm2,\ldots .
\label{eq:wavenumber_alias}
\end{equation}

Consequently, spectral weight near
$k_{\mathrm{app}}\simeq0$ indicates a small inter-probe phase
difference modulo $2\pi$, but does not uniquely establish a true
long-wavelength mode with $k_{\mathrm{true}}\simeq0$.

With the sign convention adopted above, positive $k_{\mathrm{app}}$
corresponds to propagation from Probe~2 toward Probe~1, i.e. toward the
keeper, whereas negative $k_{\mathrm{app}}$ corresponds to downstream
propagation. For each segment, the phase-derived apparent wavenumber at
each frequency was assigned to the corresponding $k_{\mathrm{app}}$ bin
and weighted by the cross-spectral magnitude $|P_{21}(f)|$. These
weighted contributions were accumulated over all segments to construct
the $f$--$k_{\mathrm{app}}$ maps.

The acquisition and segmentation parameters are summarized in
Table~\ref{tab:beall_parameters}.

\begin{table}[htbp]
\caption{Parameters used for the two-probe frequency--wavenumber
analysis.}
\label{tab:beall_parameters}
\centering
\renewcommand{\arraystretch}{1.15}
\begin{tabular}{l c}
\hline
Quantity & Value \\
\hline
Probe separation $\Delta x$ & 3.0 mm \\
Sampling frequency $f_s$ & 1.0 GHz \\
Record duration & 10.0 ms \\
FFT length & 32768 \\
Overlap & 50\% \\
Overlapping segments & 609 \\
Frequency-bin spacing & 30.52 kHz \\
\hline
\end{tabular}
\end{table}

Because the cross-spectral phase is restricted to $(-\pi,\pi]$, the
principal phase-derived wavenumber interval is

\begin{equation}
-\frac{\pi}{\Delta x}
<
k_{\mathrm{app}}
\leq
\frac{\pi}{\Delta x}.
\label{eq:principal_k_interval}
\end{equation}

For $\Delta x=3.0$~mm, the corresponding spatial-Nyquist boundary is
$
k_{\mathrm{N}}
={\pi}/{\Delta x}
=
1.047\times10^3~\mathrm{rad\,m^{-1}}.
$
Accordingly, only the principal interval
$|k_{\mathrm{app}}|\leq1000~\mathrm{rad\,m^{-1}}$ is displayed and
interpreted in the experimental maps. This principal interval does not
imply that all contributing fluctuations satisfy
$|k_{\mathrm{true}}|\leq k_{\mathrm{N}}$; shorter-wavelength
components can be mapped into the same interval through phase wrapping.

For reference, the maps also show the $n=0$ cold-ion, zero-drift
acoustic-speed relation within the principal apparent-wavenumber
interval,

\begin{equation}
f_0(k_{\mathrm{app}})
=
\frac{|k_{\mathrm{app}}|}{2\pi}
\sqrt{\frac{eT_e}{m_i}},
\label{eq:acoustic_reference}
\end{equation}

evaluated using the measured electron temperature. 
Because the ion drift velocity was not independently measured,
Eq.~\ref{eq:acoustic_reference} is used only as a cold-ion,
zero-drift acoustic-speed reference and not as a quantitative
laboratory-frame ion-acoustic dispersion relation. For the measured
electron-temperature range, this reference reaches the
spatial-Nyquist boundary at approximately 0.28--0.32~MHz. This
frequency is therefore not a hard upper-frequency limit for
ion-acoustic activity: a finite ion drift changes the laboratory-frame
phase velocity, while shorter-wavelength fluctuations may also be
aliased into the principal $k_{\mathrm{app}}$ interval. The present
two-probe measurement consequently does not exclude shorter-wavelength,
off-axis, intermittent, or spatially localized ion-acoustic activity.

% ============================================================
\subsection{PIC simulation}
\label{sec:pic_simulation}

A two-dimensional axisymmetric electrostatic PIC framework based on
Ref.~\cite{zhao2025generation} was used to generate the self-consistent
particle and electric-field histories analyzed in the present work. The
simulation was implemented in WarpX \cite{fedeli2022pushing}. The
present calculation retains the electrostatic 2D--RZ kinetic formulation
and numerical strategy developed in Ref.~\cite{zhao2025generation}, but
uses a modified computational configuration, including different domain
dimensions, electrostatic boundary potentials, injection geometry,
injected electron current, and neutral-density length scale.
Accordingly, the present PIC case is distinct from the numerical case
reported in Ref.~\cite{zhao2025generation}.

The simulation was constructed using characteristic potential,
temperature, density, and geometric scales relevant to the experiment
rather than as a point-by-point reconstruction of a particular
experimental operating condition. The configuration and numerical
parameters used here are therefore reported explicitly below. Details of
the underlying WarpX PIC implementation and collision framework are
described in Ref.~\cite{zhao2025generation}.

The computational domain, shown in
Fig.~\ref{fig:cathode_experiment}, extends
$16~\mathrm{mm}\times16~\mathrm{mm}$ in the radial and axial
directions. Dirichlet electrostatic boundary conditions were imposed at
the cathode-side and anode boundaries with
$\phi_0=15$~V and $\phi_a=45$~V, respectively, while a Neumann
condition was applied at the outer radial boundary. All three boundaries
were absorbing for particles.

Electrons and Xe$^+$ ions were injected through a circular inlet using
drifting Maxwellian velocity distributions. The electron drift velocity
was determined from a prescribed electron-current magnitude of 1.0~A,
the reference electron density
$n_{e,\mathrm{ref}}=1.0\times10^{18}~\mathrm{m^{-3}}$, and the
1.0~mm inlet radius, giving
$v_{de}=1.98\times10^6~\mathrm{m\,s^{-1}}$.
The Xe$^+$ drift velocity was prescribed independently as
$v_{di}=1.0\times10^3~\mathrm{m\,s^{-1}}$ and was not derived from
either the electron-current condition or
$n_{e,\mathrm{ref}}$.
At each time step, 400 macroparticles of each species were injected
with the same macroparticle weight,
$w_0\simeq5.5\times10^4$.
The equal-number loading and the electron and ion drift velocities are
therefore independent numerical inlet prescriptions.
The principal injection and numerical parameters are listed in
Table~\ref{simulation-parameters}.

Electron--neutral collisions were treated using a Monte Carlo collision
(MCC) model including elastic scattering, excitation, and ionization.
Ion--neutral charge-exchange collisions were not included in the
present self-consistent PIC calculation. The neutral xenon background
was prescribed rather than evolved self-consistently, with

\begin{equation}
n_a(r,z)
=
n_{a0}
\left(
1+
\frac{\sqrt{r^2+z^2}}{L_0}
\right)^{-2},
\label{eq:neutral_density}
\end{equation}

where $n_{a0}=1.0\times10^{19}~\mathrm{m^{-3}}$ and
$L_0=1.6~\mathrm{cm}$. The neutral temperature was fixed at 0.5~eV.

The simulation was advanced to
$t_{\mathrm{sim}}=17.7259~\mu\mathrm{s}$.
For the qualitative spectral comparison with the experimental
potential-fluctuation proxy, Fourier-amplitude spectra were calculated
as $|\mathrm{FFT}[\widetilde{\phi}(t)]|/N$, where $N$ is the number of
samples. The experimental and simulated spectra were normalized
independently by their respective maxima before comparison.

The simulated potential fluctuations were also characterized in
frequency--wavenumber space using the electrostatic potential sampled
along the axial direction at the fixed radial grid location $N_r=5$.
The axial $f$--$k$ spectrum was obtained from a spatiotemporal Fourier
analysis of $\widetilde{\phi}(z,t)$, with the acoustic-speed reference
evaluated using $T_e=5.0$~eV.

The time-dependent radial and axial electric fields extracted from the
PIC calculation were subsequently used in the particle-resolved
analyses described in Sec.~\ref{sec:particle_tracking}.

\begin{table}[htbp]
\caption{Parameters used in the hollow-cathode plume PIC simulation.}
\label{simulation-parameters}
\centering
\renewcommand{\arraystretch}{1.12}
\resizebox{\columnwidth}{!}{%
\begin{tabular}{l c c}
\hline
Quantity & Symbol & Value \\
\hline
Reference electron density
    & $n_{e,\mathrm{ref}}$ & $1.0\times10^{18}$ m$^{-3}$ \\
Boundary potentials
    & $\phi_0,\phi_a$ & 15 V, 45 V \\
Injection radius
    & $r_0$ & 1.0 mm \\
Injected electron current
    & $I_0$ & 1.0 A \\
Electron temperature
    & $T_e$ & 5.0 eV \\
Ion temperature
    & $T_i$ & 0.5 eV \\
Electron drift velocity
    & $v_{de}$ & $1.98\times10^6$ m s$^{-1}$ \\
Prescribed Ion drift velocity
    & $v_{di}$ & $1.0\times10^3$ m s$^{-1}$ \\
Neutral-density coefficient
    & $n_{a0}$ & $1.0\times10^{19}$ m$^{-3}$ \\
Neutral length scale
    & $L_0$ & 1.6 cm \\
Neutral temperature
    & $T_a$ & 0.5 eV \\
Domain size
    & $L_r\times L_z$ & 16 mm $\times$ 16 mm \\
Cell size
    & $\Delta r=\Delta z$ & $1.0\times10^{-5}$ m \\
Debye length
    & $\lambda_D$ & $1.66\times10^{-5}$ m \\
Time step
    & $\Delta t$ & $3.55\times10^{-12}$ s \\
Injected macroparticles/step/species
    & $N_p$ & 400 \\
Macroparticle weight
    & $w_0$ & $\approx5.5\times10^4$ \\
Total duration
    & $t_{\mathrm{sim}}$ & 17.7259 $\mu$s \\
\hline
\end{tabular}}
\end{table}

% ============================================================
\subsection{Particle-resolved analysis}
\label{sec:particle_tracking}

A kinetic-energy threshold of 50~eV was used throughout the
particle-resolved analysis to define the energetic-ion population. This
threshold lies well above the energy scale of the imposed inlet-ion
distribution.

Two statistical ensembles are distinguished. For the self-consistent
PIC plasma, the instantaneous energetic-ion fraction is defined as

\begin{equation}
f_{\mathrm{SC}}(t)
=
\frac{
N_{\mathrm{SC}}\!\left(K_i(t)>50~\mathrm{eV}\right)
}{
N_{\mathrm{SC}}(t)
}.
\label{eq:selfconsistent_fraction}
\end{equation}

For a passive birth cohort generated from source model $s$ and
propagated through field model $m$, the conditional energetic fraction
is defined as

\begin{equation}
f_{\mathrm{passive}}^{(s,m)}
=
\frac{
N_{\mathrm{passive},>50}^{(s,m)}
}{
N_{\mathrm{passive},0}^{(s)}
}.
\label{eq:passive_fraction}
\end{equation}

For passive particles, the final kinetic energy $K_f$ is evaluated at
absorbing-boundary crossing or, for particles remaining in the domain,
at the end of the propagation interval. Thus,
$f_{\mathrm{SC}}$ is an instantaneous self-consistent population
fraction, whereas $f_{\mathrm{passive}}^{(s,m)}$ is a conditional
response of a prescribed birth cohort; the two quantities are not
directly comparable.

Passive ions were propagated through the stored PIC electric field
without depositing charge or modifying the prescribed field evolution.
The complete stored history contained 501 snapshots over
$0\leq t\leq17.7259~\mu\mathrm{s}$ with a spacing of 35.4518~ns.
Unless otherwise stated, particles were initialized at
$t_0=3.7259~\mu\mathrm{s}$ and propagated to
$t_1=8.7259~\mu\mathrm{s}$ using a particle time step of 0.1~ns.
This field interval is indicated in
Fig.~\ref{fig:pic_temporal_evolution}.

The electric field was evaluated by linear interpolation between stored
snapshots in time and bilinear interpolation on the $r$--$z$ mesh in
space. 
Passive propagation was collisionless; no collisional processes were
applied, and particles were removed upon crossing an absorbing boundary.

The passive-ion equations of motion were

\begin{equation}
\frac{d\mathbf{x}_p}{dt}
=
\mathbf{v}_p,
\qquad
\frac{d\mathbf{v}_p}{dt}
=
\frac{q_i}{m_i}
\mathbf{E}(\mathbf{x}_p,t).
\label{eq:passive_motion}
\end{equation}

% ------------------------------------------------------------
\subsubsection{Source and birth-position analysis}
\label{sec:source_birth_method}
% ------------------------------------------------------------

Self-consistent PIC ions crossing either the downstream boundary
$z_{\max}$ or the outer radial boundary $r_{\max}$ with
boundary-crossing kinetic energies above 50~eV were combined into an
energetic-outflow subset and classified according to their
particle-source flags. Ions carrying the imposed injection flag were
identified as inlet-injected particles, whereas ions created inside the
computational domain without this flag were identified as
ionization-born particles.

The ionization-born fraction of the energetic outward population was

\begin{equation}
f_{\mathrm{ion}\mid E,\mathrm{out}}
=
\frac{
N_{\mathrm{ion},>50}^{\mathrm{out}}
}{
N_{>50}^{\mathrm{out}}
}.
\label{eq:conditional_source_fraction}
\end{equation}

This quantity describes the source composition of the energetic outflow,
rather than the probability for an ionization-born ion to become
energetic.

Two source-related passive-particle controls were performed. First,
$5\times10^6$ Xe$^+$ ions were launched through the inlet using the
same initial velocity distribution as the injected ions in the
self-consistent PIC simulation. This calculation tested whether the
inlet population could access trajectories reaching the downstream or
radial boundaries.

Second, sensitivity to ion birth position was examined using three
spatial source models:
(i) an empirical kernel-density estimate (KDE) constructed from
ionization events recorded in the self-consistent PIC simulation,
(ii) a distribution uniform in physical volume, and
(iii) a neutral-density-weighted distribution in the $r$--$z$
coordinate plane based on Eq.~\ref{eq:neutral_density}.

In the axisymmetric geometry, uniform physical-volume sampling accounts
for the cylindrical volume element $r\,dr\,dz$. For the
neutral-density-weighted model, the sampling probability per unit
$r$--$z$ coordinate area was proportional to $n_a(r,z)$; no additional
cylindrical-volume weighting was applied.

Each birth-position cohort contained $10^7$ Xe$^+$ ions with an initial
kinetic energy of 0.5~eV. Initial velocity directions were sampled
isotropically using the same rule for all three source models. All
cohorts were initialized at the same time and propagated through the
same full time-dependent PIC electric-field history. The prescribed
spatial birth-position distribution was therefore the only controlled
difference among the three calculations.

% ------------------------------------------------------------
\subsubsection{Electric-field model controls}
\label{sec:field_model_controls}
% ------------------------------------------------------------

The role of the electric-field time history was examined using a matched
empirical-KDE realization containing $10^6$ Xe$^+$ ions with 0.5~eV
initial kinetic energy and isotropic initial velocity directions.
Exactly the same initial particle positions and velocities were used in
all field controls.

Four prescribed electric-field models were compared over the
$3.7259$--$8.7259~\mu\mathrm{s}$ interval. The reference case retained
the full time-dependent PIC field,
$\mathbf{E}_{\mathrm{PIC}}(r,z,t)$. The time-averaged field was

\begin{equation}
\overline{\mathbf{E}}(r,z)
=
\frac{1}{N_t}
\sum_{j=1}^{N_t}
\mathbf{E}_{\mathrm{PIC}}(r,z,t_j),
\label{eq:mean_field}
\end{equation}

where the average was taken over the 141 stored snapshots in the
propagation interval.

A frozen-field control used the stored PIC field snapshot closest to the
midpoint of the interval,
$t_f=6.2395~\mu\mathrm{s}$,
and held this spatial field fixed throughout the particle propagation.
The fluctuation-only field was defined as

\begin{equation}
\delta\mathbf{E}(r,z,t)
=
\mathbf{E}_{\mathrm{PIC}}(r,z,t)
-
\overline{\mathbf{E}}(r,z).
\label{eq:fluctuation_field}
\end{equation}

The four field models are interpreted as matched sensitivity tests of
the electric-field time history rather than as an additive decomposition
of the physical ion-energy gain.

\subsubsection{Particle work analysis}
\label{sec:work_method}

The trajectory-resolved work analysis used the empirical-KDE cohort
from the birth-position analysis, containing $10^7$ passive Xe$^+$ ions
propagated through the full time-dependent PIC electric field. The
particle-level energy transfer along each trajectory was quantified from

\begin{equation}
\frac{dK_i}{dt}
=
q_i\mathbf{E}\cdot\mathbf{v}_p
=
q_i(E_rv_r+E_zv_z),
\label{eq:particle_power}
\end{equation}
where $v_r$ and $v_z$ are the radial and axial ion-velocity components.

The accumulated radial and axial work were
$W_r(t)
=
\int_{t_0}^{t}
q_iE_rv_r\,dt'$
and
$W_z(t)
=
\int_{t_0}^{t}
q_iE_zv_z\,dt'$,
with
$W_{\mathrm{net}}(t)
=
W_r(t)+W_z(t)$.
For the collisionless passive trajectories,

\begin{equation}
K_i(t)-K_i(t_0)
=
W_{\mathrm{net}}(t),
\label{eq:particle_work_energy}
\end{equation}

which was also used as a consistency check on the trajectory integration
and field interpolation.

Spatial positive-work maps were constructed by accumulating only
trajectory increments satisfying
$q_i\mathbf{E}\cdot\mathbf{v}_p>0$.
For each escape family, the accumulated positive work in each spatial
bin was normalized by the number of sampled ions in that family, giving
the positive work per sampled ion per spatial bin. These maps identify
regions contributing positive kinetic-energy transfer and do not
represent the net work accumulated over the complete trajectory. The
accumulated $W_r$, $W_z$, and $W_{\mathrm{net}}$ were grouped according
to final ion energy and escape boundary, while particle residence times
were analyzed in parallel.

To illustrate representative single-particle energy-gain pathways,
particles were treated separately according to their escape boundary,
$z_{\min}$, $z_{\max}$, or $r_{\max}$. For each escape family,
particles with $50\leq K_f\leq250$~eV were first classified according
to the signed radial fraction of the net accumulated work,
$f_r={W_r}/{(W_r+W_z)}$.
Particles with $f_r<0.25$, $0.25\leq f_r\leq0.75$, and $f_r>0.75$
were classified as axial-work dominated, mixed, and radial-work
dominated, respectively. Representative particles were selected from
the most populated (dominant) work class of each escape family and
restricted to the central 40th--60th percentile range of the
final-energy distribution.

Within the central-energy subset, representative particles were
selected algorithmically as robust medoids of the dominant
directional-work class using particle-energy, residence-time,
directional-work, and pathway-history characteristics. The selected
pathway in each escape family therefore corresponds to an actual
simulated particle rather than an averaged or manually chosen
trajectory.

% Within the central-energy subset, $K_f$, residence time $\tau$, and
% $f_r$ were standardized using the median and interquartile range (IQR),
% and the 12 real particles closest to this robust center were retained
% for each escape family. These candidates were replayed through the same
% time-dependent PIC field to obtain the additional pathway descriptors
% $\eta_-=|W^-|/W^+$, $t_{50}/\tau$, $z_W/L_z$, and $r_W/R$.
% The final pathway feature vector was therefore
% $\mathbf{u}=[K_f,\tau,f_r,\eta_-,t_{50}/\tau,z_W/L_z,r_W/R]$.
% Before the medoid calculation, candidates for which any feature deviated
% by more than five IQRs from the candidate median were excluded. After
% median--IQR normalization, the representative particle was defined as
% the actual candidate minimizing the total Euclidean distance to the
% other retained candidates in this normalized feature space,
% $
% i_{\mathrm{medoid}}
% =
% \arg\min_i
% \sum_j
% \left\|
% \mathbf{u}_i-\mathbf{u}_j
% \right\|_2 .
% $
% Thus, the representative pathway corresponds to a real simulated
% particle rather than an averaged or reconstructed trajectory.

\section{Results}
\label{sec:results}

\subsection{Energetic ions and plume fluctuations in the low-current regime}
\label{sec:experimental_observation}

\begin{figure}[htbp]
    \centering
    \includegraphics[width=0.95\linewidth]
    {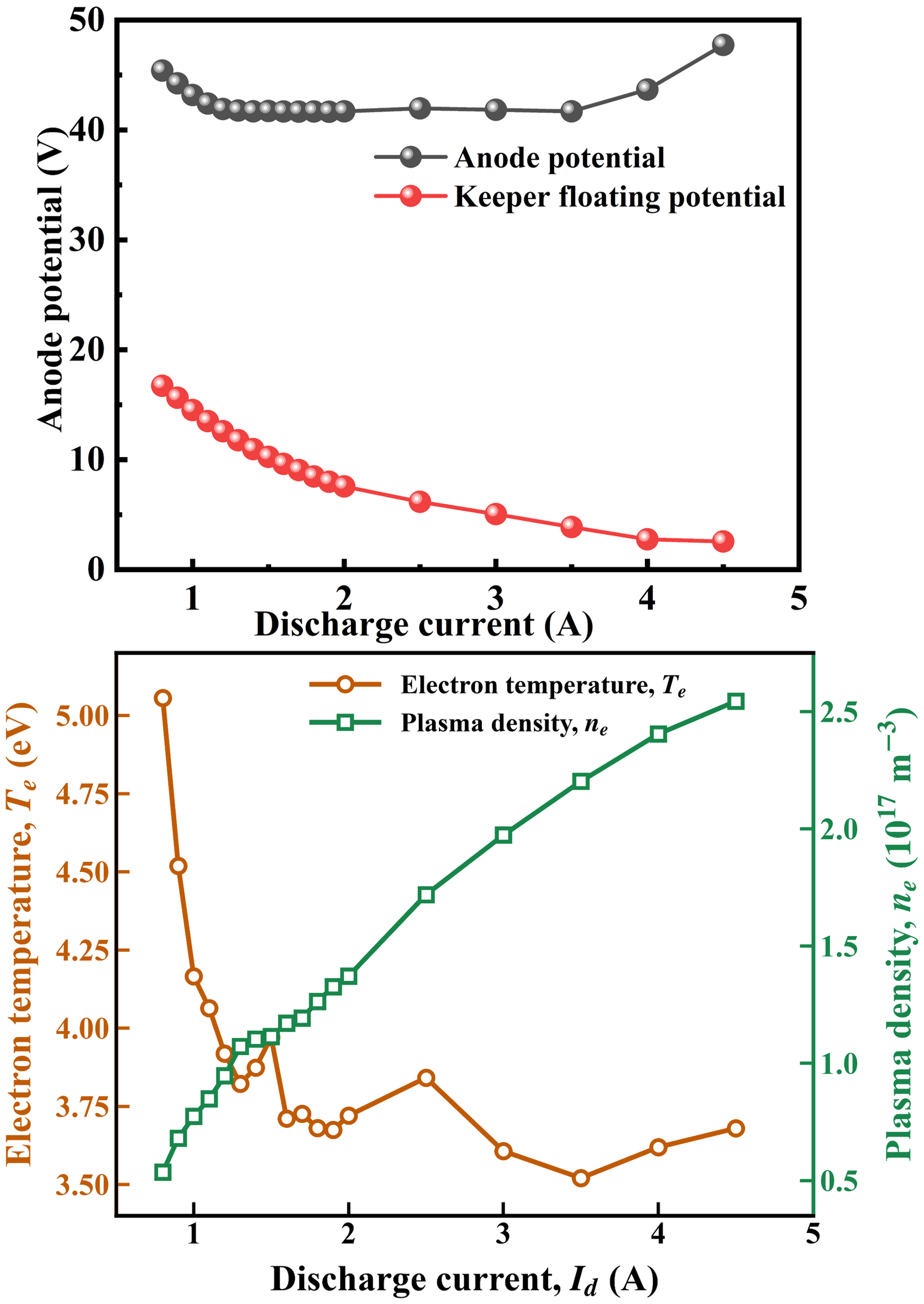}
    \caption{Experimentally measured hollow-cathode plume parameters as
    a function of discharge current: (a) anode potential and keeper
    floating potential, (b) electron temperature and plasma density.}
    \label{fig:experimental_data}
\end{figure}

The experimental measurements were first examined to establish the
plume conditions over the investigated low-current operating range.
The measured discharge and plasma parameters are summarized in
figure~\ref{fig:experimental_data}.

The cathode was operated with xenon at a fixed flow rate of
1.5~sccm while the discharge current was varied from 0.8 to 3.5~A.
As shown in figure~\ref{fig:experimental_data}(a), the anode potential
remains approximately within 42--45~V over most of the operating range
and increases at the highest current, whereas the keeper floating
potential decreases with increasing current. The electron temperature
decreases from approximately 5~eV to about 3.8~eV and then varies only
weakly, whereas the plasma density increases continuously with discharge
current [figure~\ref{fig:experimental_data}(b)].

\begin{figure*}[htbp]
    \centering
    \includegraphics[width=0.32\linewidth]
    {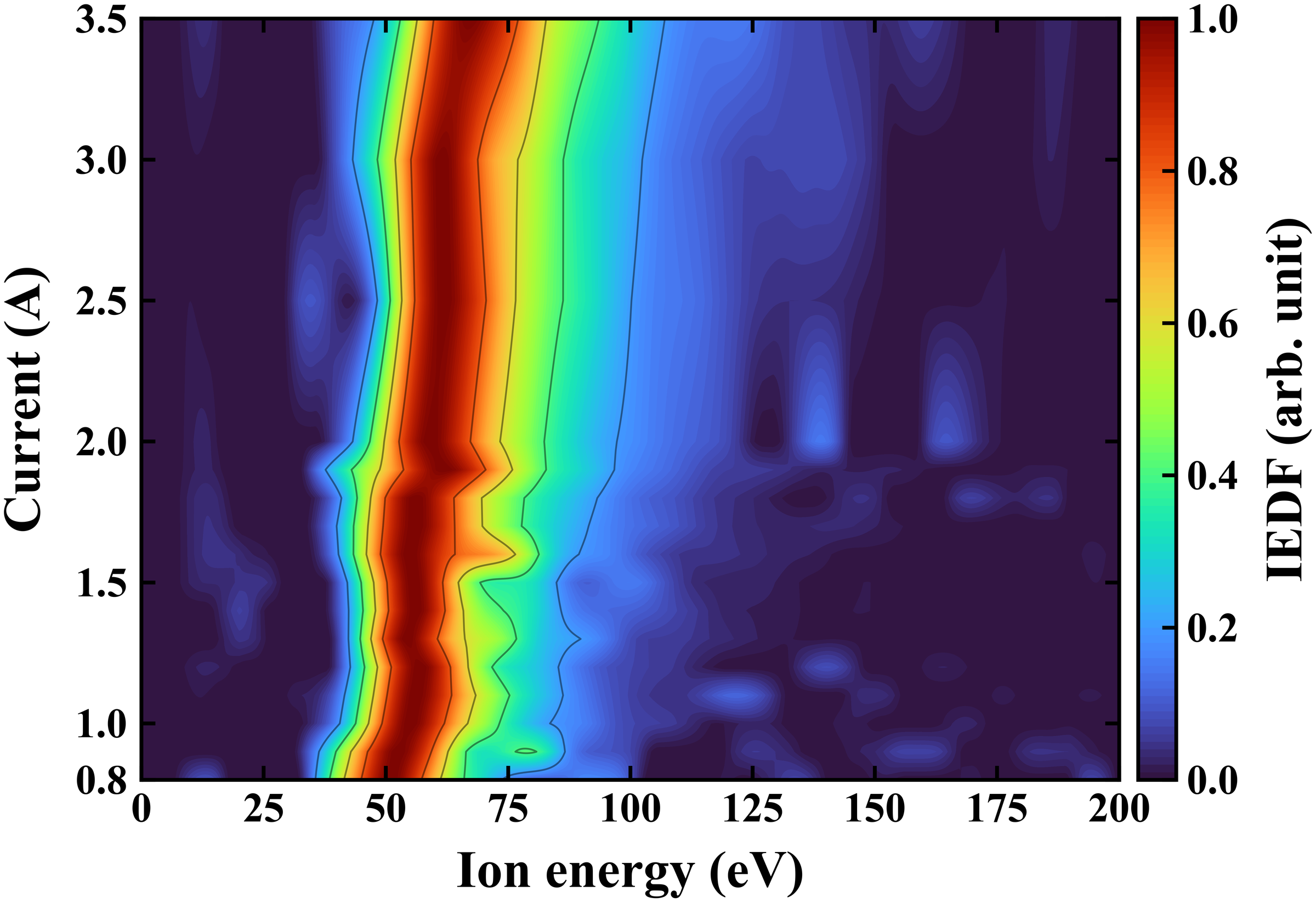}
    \includegraphics[width=0.32\linewidth]
    {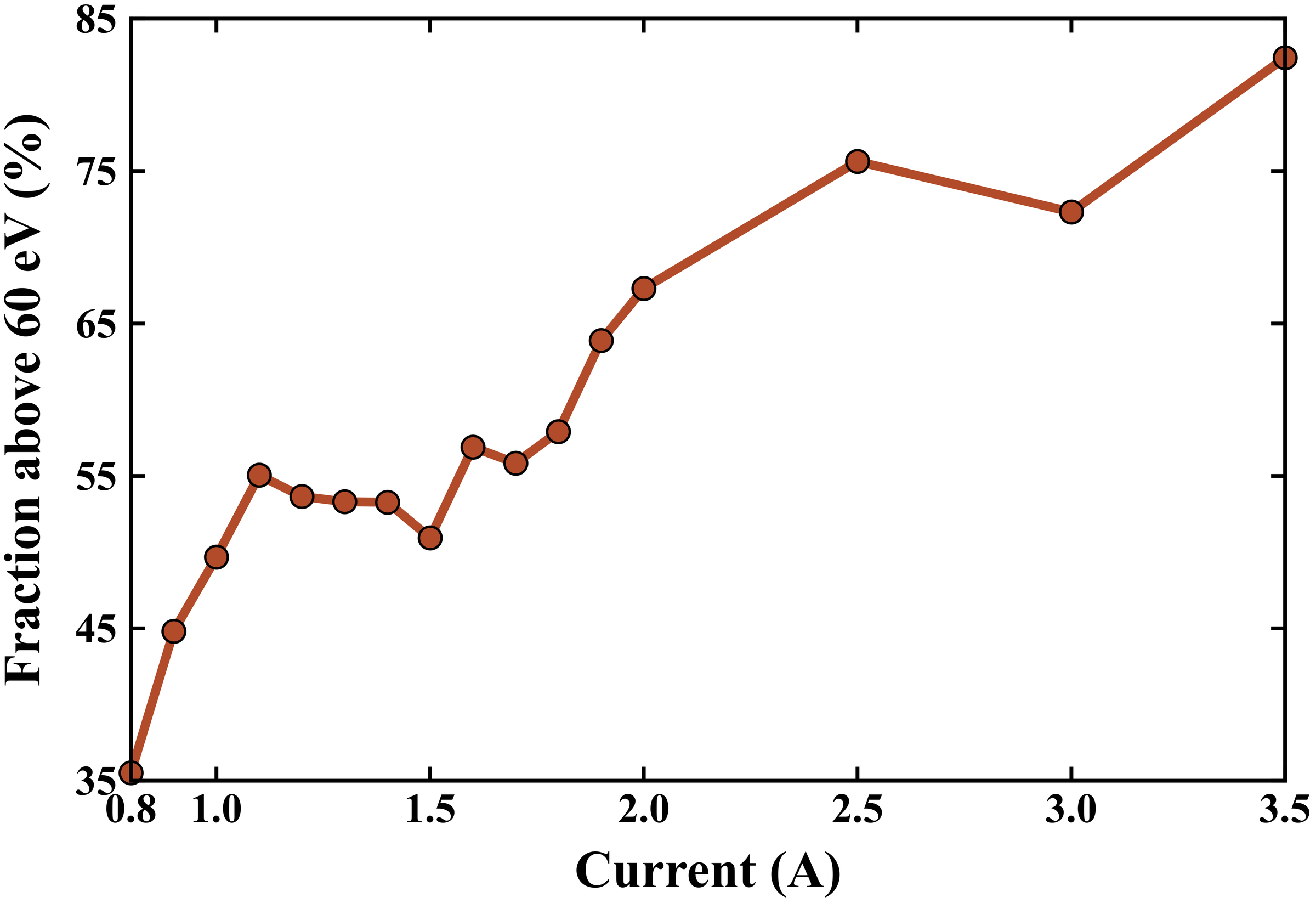}
    \includegraphics[width=0.32\linewidth]
    {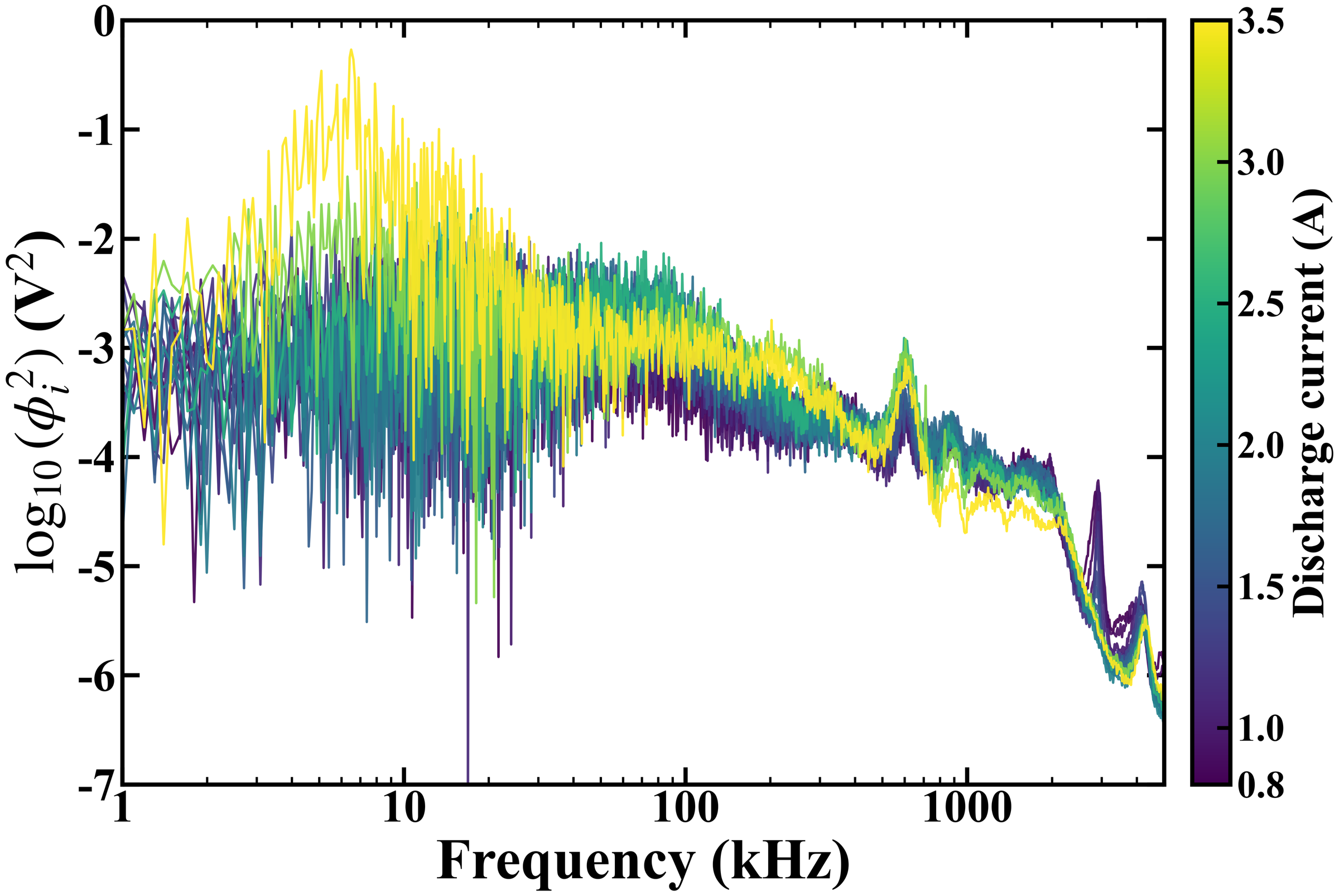}
    \caption{Experimental energetic-ion and fluctuation characteristics
    as functions of discharge current. (a) Xe$^+$-equivalent ion-energy distributions inferred from the retarding potential analyzer. (b) Fraction of the processed RPA-derived distribution above a Xe$^+$-equivalent energy of 60~eV. (c) Logarithmic power
    spectra of the measured ion-saturation-current fluctuations.}
    \label{ion-energy}
\end{figure*}

Figure~\ref{ion-energy} summarizes the measured energetic-ion and
fluctuation characteristics over the investigated discharge-current
range. The RPA distributions exhibit a substantial energetic-ion
population throughout the measurements
[figure~\ref{ion-energy}(a)]. The dominant ion-energy feature shifts
from approximately 50--60~eV at lower discharge current toward about
70~eV at higher current. Meanwhile, the fraction of the processed RPA distribution above a
Xe$^+$-equivalent energy of 60~eV increases from approximately 40\%
to nearly 80\% [figure~\ref{ion-energy}(b)]. Energetic ions are therefore present throughout the investigated low-current regime and become increasingly
prominent with increasing discharge current.

The corresponding ion-saturation-current spectra contain broadband
fluctuations extending from the tens-of-kilohertz range into the
megahertz range [figure~\ref{ion-energy}(c)]. Broadband spectral
content alone, however, does not establish the presence of a propagating
ion-acoustic mode. To examine whether the observed fluctuations contain
a resolved acoustic-like propagation signature, the synchronized probe
signals were therefore analyzed in frequency--wavenumber space using
the two-probe method described in section~\ref{sec:fk_method}.

\begin{figure*}[htbp]
    \centering
    \includegraphics[width=0.98\linewidth]
    {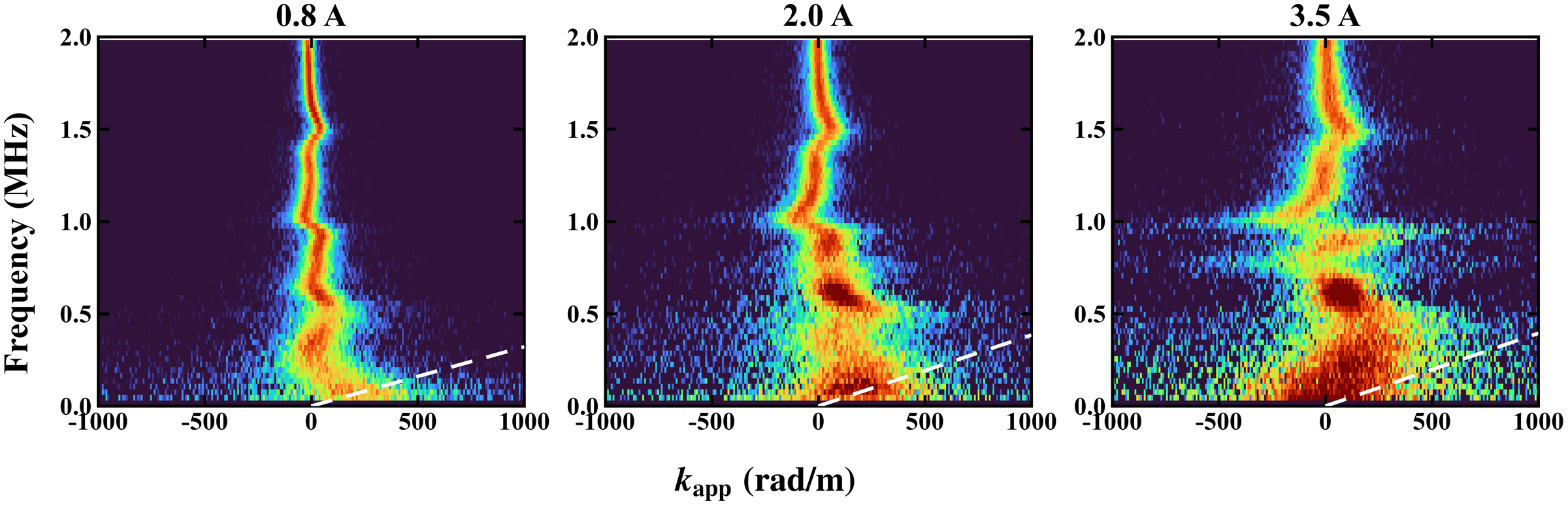}
    \caption{Representative ion-saturation-current
cross-spectral-magnitude-weighted, two-point phase-derived
frequency--wavenumber maps at discharge currents of
(a) 0.8~A, (b) 2.0~A, and (c) 3.5~A.
The color scale represents the accumulated cross-spectral magnitude
$|P_{21}(f)|$ in each $f$--$k_{\mathrm{app}}$ bin.
The dashed curves denote the $n=0$ cold-ion, zero-drift
acoustic-speed reference evaluated using the measured electron
temperature. 
% Only the principal apparent-wavenumber interval
% $|k_{\mathrm{app}}|\leq1000~\mathrm{rad\,m^{-1}}$ is shown.
% Because $k_{\mathrm{app}}$ is inferred from the wrapped two-point
% cross phase, spectral weight near $k_{\mathrm{app}}=0$ does not
% uniquely imply $k_{\mathrm{true}}=0$.
}
    \label{dispersion_relation}
\end{figure*}

Representative two-point phase-derived frequency--wavenumber maps at
discharge currents of 0.8, 2.0, and 3.5~A are shown in
figure~\ref{dispersion_relation}, spanning the low-, intermediate-, and
high-current portions of the investigated range. The horizontal
coordinate is the apparent wavenumber $k_{\mathrm{app}}$ defined by
Eq.~\ref{eq:beall_wavenumber}, and the dashed curves denote the
$n=0$ cold-ion, zero-drift acoustic-speed reference of
Eq.~\ref{eq:acoustic_reference}.

The representative maps evolve from broadband cross-spectral weight
concentrated predominantly near small
$|k_{\mathrm{app}}|$ at lower discharge current toward more pronounced
horizontal and oblique features at higher current. Importantly, the
near-$k_{\mathrm{app}}=0$ spectral weight should not be interpreted as
a distinct physical $k_{\mathrm{true}}=0$ dispersion branch.
Because $k_{\mathrm{app}}$ is inferred from the principal two-point
cross phase, this feature indicates that the two probe signals are
nearly in phase modulo $2\pi$ over a broad frequency range.
Large-scale spatially coherent plume fluctuations, shorter-wavelength
components aliased into the principal interval, and superposed
propagating components may all contribute to such spectral weight.

Although portions of the spectra overlap the acoustic-speed reference,
no continuous, approximately linear ridge persists over a finite
$f$--$k_{\mathrm{app}}$ interval that can be unambiguously identified
as a resolved ion-acoustic branch. The remaining measured current
conditions likewise do not exhibit a clearly resolved continuous
acoustic-like branch within the principal apparent-wavenumber interval.

The experiment therefore establishes the coexistence of a substantial
energetic-ion population and broadband, structured plume fluctuations.
The present two-point measurements do not establish the absence of
ion-acoustic activity; rather, they show that no continuous
ion-acoustic branch can be unambiguously resolved within the available
principal apparent-wavenumber interval. Shorter-wavelength activity
folded by spatial aliasing, as well as intermittent, off-axis, or
spatially localized ion-acoustic activity, remains possible.
Consequently, the available collective measurements do not provide an
unambiguous modal attribution for the energetic-ion population. The
particle-level formation pathway is examined using the kinetic
simulation below.

\subsection{Self-consistent plume dynamics and energetic-ion population}
\label{sec:PIC_formation}

The modified representative PIC case described in
section~\ref{sec:pic_simulation} was first characterized at the
collective level before its particle and electric-field histories were
used for the particle-resolved analysis. The simulation develops a
time-dependent kinetic plume containing broadband electrostatic
fluctuations together with a nonthermal energetic-ion population. These
collective characteristics establish the self-consistent kinetic
environment in which the subsequent source-, trajectory-, field-, and
work-resolved analyses are performed. Because the simulation is not a
point-by-point reconstruction of a specific experimental operating
condition, the experimental--numerical comparisons below are interpreted
qualitatively.

\begin{figure*}[htbp]
    \centering
    \includegraphics[width=0.95\linewidth]{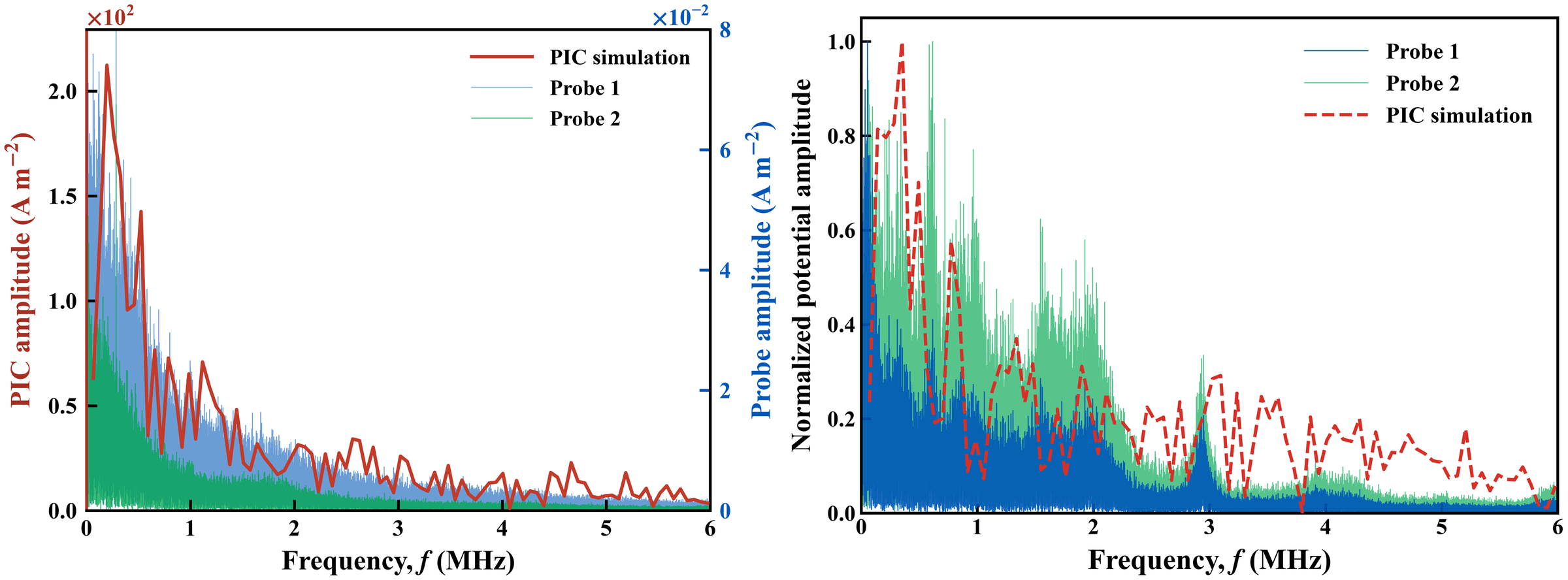}
    \caption{Qualitative spectral comparison between the experiment and
    the present representative PIC plume. (a) Comparison of the simulated
    anode-current spectrum and the experimentally measured probe-current
    spectra. (b) Comparison of the potential-fluctuation proxies
    inferred from Probe~1 and Probe~2 with the local electrostatic-potential
    fluctuation obtained directly from the PIC simulation. In panel (b),
    each spectrum is normalized independently by its own maximum to
    emphasize characteristic frequency content rather than absolute
    spectral amplitude. These comparisons are intended to assess
    broadband spectral behavior qualitatively rather than to establish
    quantitative agreement between different observables.}
    \label{fig:exp_pic_spectrum_comparison}
\end{figure*}

The simulated anode-current spectrum exhibits broadband frequency
content, as shown together with the experimentally measured
probe-current spectra in figure~\ref{fig:exp_pic_spectrum_comparison}.
The simulated plume also develops broadband local potential
fluctuations. Figure~\ref{fig:exp_pic_spectrum_comparison} compares the
normalized Fourier-amplitude spectrum of the directly resolved PIC
potential fluctuation with the experimental potential-fluctuation
proxies inferred from the two probes using
Eq.~\ref{eq:potential_proxy}. Because the experimental proxies and
simulated potential differ in physical definition and sampling, each
spectrum is normalized independently and the comparison is restricted
to characteristic frequency content.

% \begin{figure}[htbp]
%     \centering
%     \includegraphics[width=0.9\linewidth]
%     {figure/Results/two_probes_pic_normalized_potential_spectra_layered.png}
%     \caption{Normalized Fourier-amplitude spectra of the
%     potential-fluctuation proxies inferred from Probe~1 and Probe~2 and
%     the local electrostatic-potential fluctuation obtained directly from
%     the PIC simulation. Each spectrum is normalized independently by its
%     own maximum to emphasize characteristic frequency content rather
%     than absolute spectral amplitude.}
%     \label{fig:potential-probe-comparison}
% \end{figure}

\begin{figure}[htbp]
    \centering
    \includegraphics[width=0.95\linewidth]
    {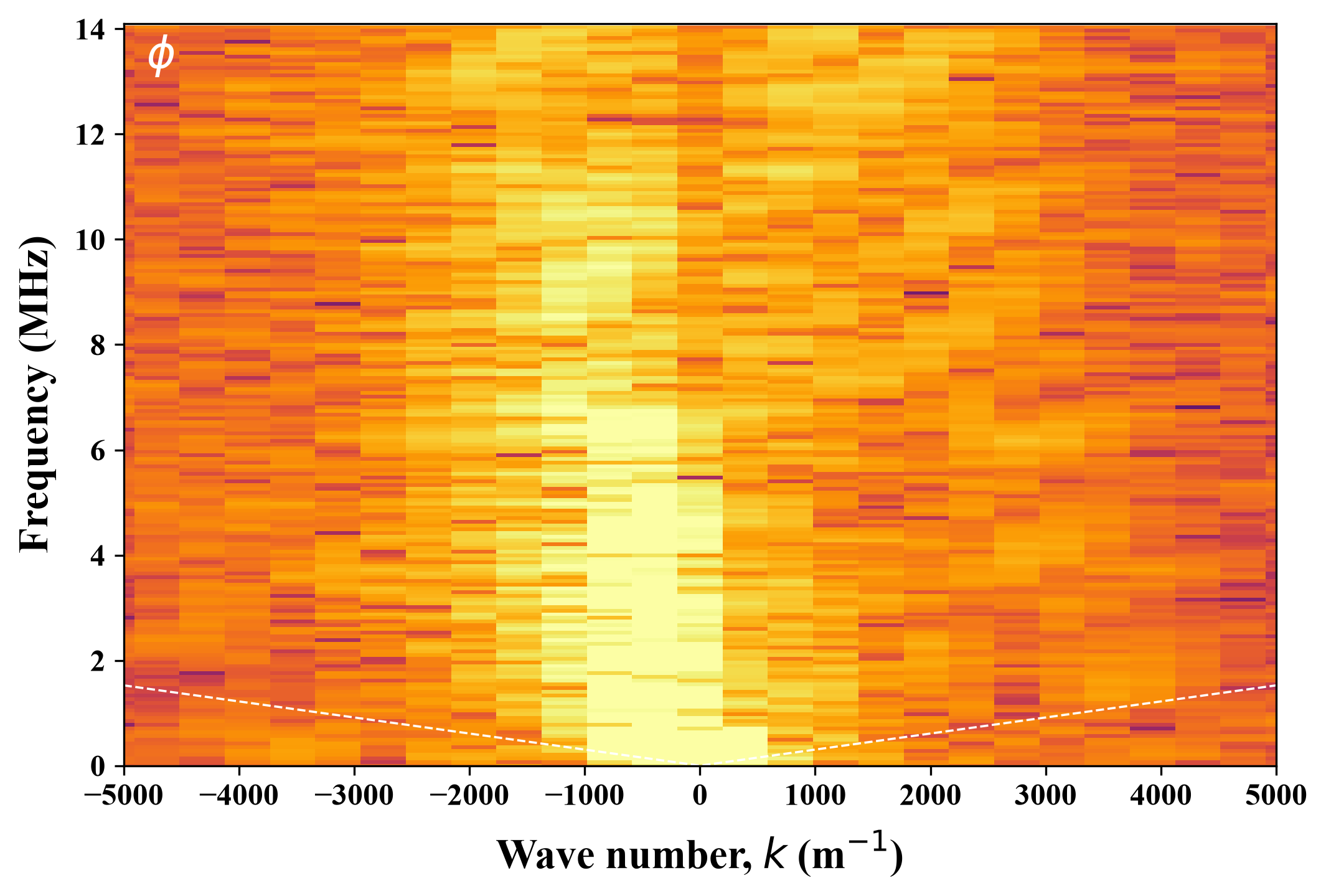}
    \caption{Frequency--wavenumber spectrum of the simulated
    electrostatic-potential fluctuation at the fixed radial grid location
    $N_r=5$. The white dashed line denotes the acoustic-speed reference,
    $f_0=|k|c_s/(2\pi)$, and is used only to assess whether a continuous
    spectral feature follows an acoustic-like propagation relation.}
    \label{fig:pic_phi_dispersion}
\end{figure}

The corresponding numerical frequency--wavenumber spectrum is shown in
figure~\ref{fig:pic_phi_dispersion}. 
The simulated spectrum contains
broadband fluctuation power, with the strongest intensity concentrated
predominantly at relatively small wavenumbers. No single continuous
ridge follows the acoustic-speed reference over a finite wavenumber
interval. The spectrum is therefore used here to characterize the
collective dynamics of the present PIC plume rather than to assign
its time-dependent electric field to a specific plasma mode.

\begin{figure*}[t]
    \centering
    \includegraphics[width=0.9\linewidth]
    {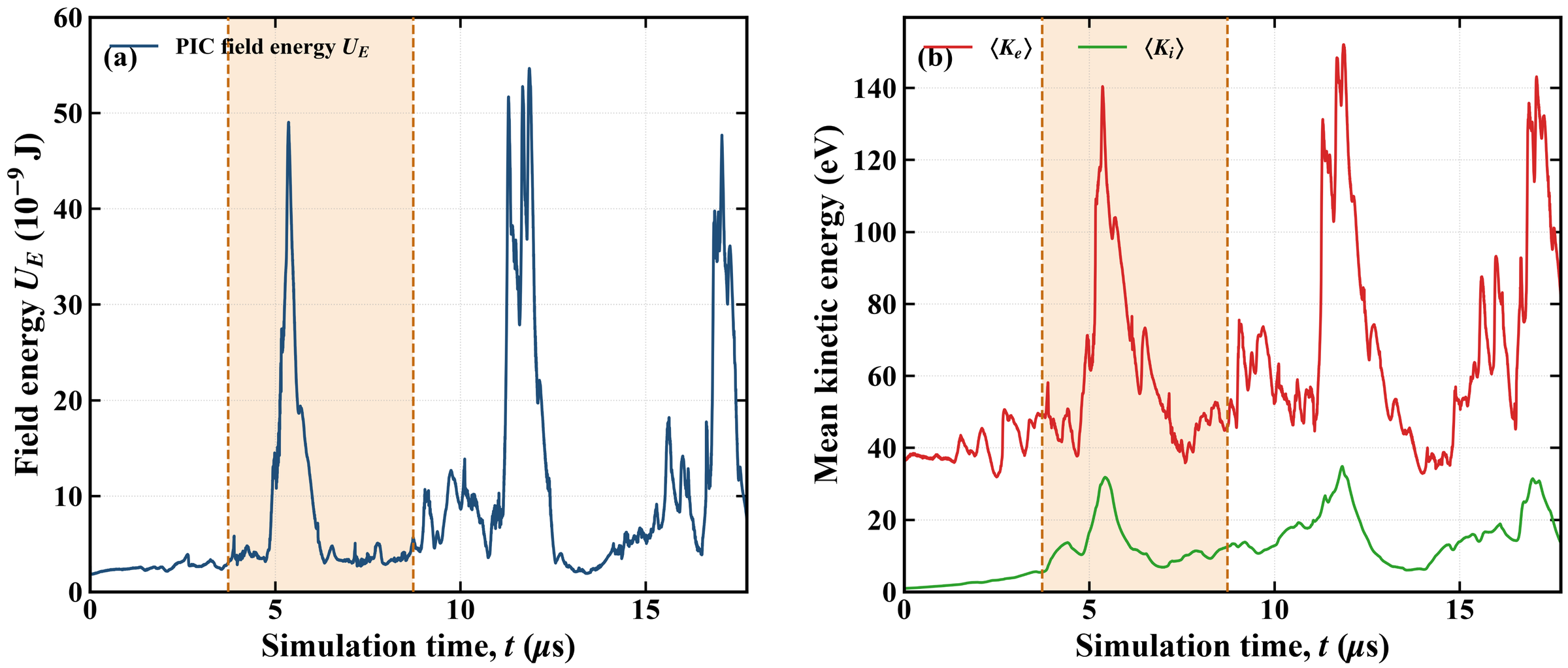}
    \caption{Temporal evolution of the self-consistent PIC plume:
    (a) electrostatic field energy $U_E$ and
    (b) mean electron and ion kinetic energies,
    $\langle K_e\rangle$ and $\langle K_i\rangle$.
    The shaded region denotes the
    $3.7259$--$8.7259~\mu\mathrm{s}$ field interval used for the
    passive-particle analyses.}
    \label{fig:pic_temporal_evolution}
\end{figure*}
The global temporal evolution of the self-consistent plume was also
examined to place the field interval used for the subsequent
particle-resolved analyses in context. As shown in
figure~\ref{fig:pic_temporal_evolution}, the
$3.7259$--$8.7259~\mu\mathrm{s}$ interval used for the subsequent
particle-resolved analyses spans appreciable temporal variation in both
the electrostatic field energy and the mean particle kinetic energies.
The corresponding particle calculations therefore probe ion dynamics
within an evolving rather than nominally steady field history.

% \begin{figure*}[t]
%     \centering
%     \includegraphics[width=1.0\linewidth]
%     {figure/Results/fig_energy_quantiles_vs_time.png}
%     \caption{Temporal evolution of electron and ion energy quantiles in
%     the self-consistent PIC simulation. The growth of the upper
%     ion-energy percentiles indicates the development of a nonthermal
%     energetic-ion population during plume evolution.}
%     \label{fig:energy_quantiles_time}
% \end{figure*}
% The development of the energetic-ion population is more directly
% resolved by the ion-energy quantiles shown in
% figure~\ref{fig:energy_quantiles_time}.
% At $t=5.32~\mu$s, the mean ion energy is approximately 29.8~eV and the
% 90th percentile reaches 67.3~eV. At the same instant, 11.87\% of all
% self-consistent PIC ions present in the simulation domain have kinetic
% energies above 50~eV, corresponding to
% $f_{\mathrm{SC}}(5.32~\mu\mathrm{s})=0.1187$ according to
% Eq.~\ref{eq:selfconsistent_fraction}.
% The energetic tail therefore emerges during the
% self-consistent kinetic evolution rather than being introduced through
% a prescribed high-energy source.

The self-consistent PIC plume also develops a substantial
energetic-ion population. At $t=5.32~\mu\mathrm{s}$, the mean ion
kinetic energy is approximately 29.8~eV, while 11.87\% of the ions
present in the simulation domain have kinetic energies above 50~eV,
corresponding to
$f_{\mathrm{SC}}(5.32~\mu\mathrm{s})=0.1187$
according to Eq.~\ref{eq:selfconsistent_fraction}.
The origin of this energetic population is examined below using
particle-source and trajectory histories.

\begin{figure}[htbp]
    \centering
    \includegraphics[width=0.9\linewidth]
    {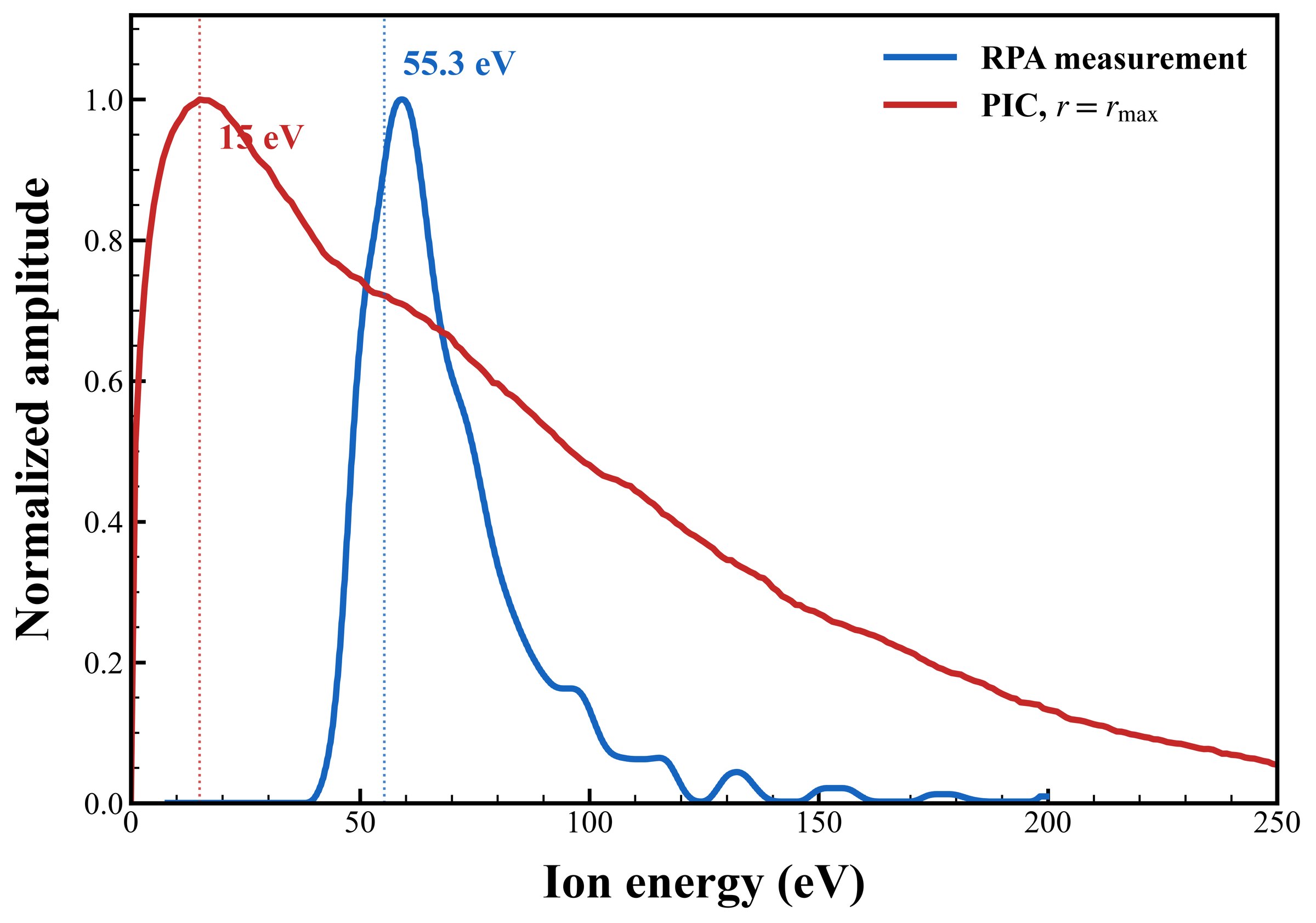}
    \caption{Normalized radial ion-energy signals obtained from the RPA
measurement and the PIC simulation. The experimental RPA samples
radially propagating ions, while the numerical distribution is
constructed from Xe$^+$ ions crossing the outer radial boundary
$r_{\max}$. The comparison is intended only as a qualitative comparison of the radial energetic-ion components rather than quantitative agreement of
the distribution shape. 
}
    \label{fig:cathode_experiment_rpa}
\end{figure}

For qualitative context, normalized ion-energy signals from the RPA and
PIC simulation are compared in
figure~\ref{fig:cathode_experiment_rpa}.
The RPA distribution exhibits a pronounced maximum near 55.3~eV and a
finite component above 100~eV, whereas the PIC population sampled at the
radial boundary extends to approximately 250~eV. 
Although the two populations correspond in their nominal radial
sampling direction, they differ in operating condition, sampling
location, analyzer acceptance, and diagnostic response. The comparison
is therefore restricted to qualitative consistency in the presence of a
substantial nonthermal radial energetic-ion component.

The PIC results therefore provide a self-consistent kinetic environment
containing broadband temporal fluctuations and a nonthermal
energetic-ion population. The following particle-resolved analysis
examines how this population is connected to source localization,
trajectory accessibility, temporal electric-field evolution, and
particle-level energy transfer.

\subsection{Particle-resolved energetic-ion pathway}
\label{sec:kinetic_origin}

The particle-resolved analysis proceeds from ion source and trajectory
accessibility to electric-field-history sensitivity and finally to
trajectory-integrated energy transfer.

\subsubsection{Ion source and trajectory accessibility}
\label{sec:ion_source}

\begin{figure}[t]
    \centering
    \includegraphics[width=1.0\linewidth]
    {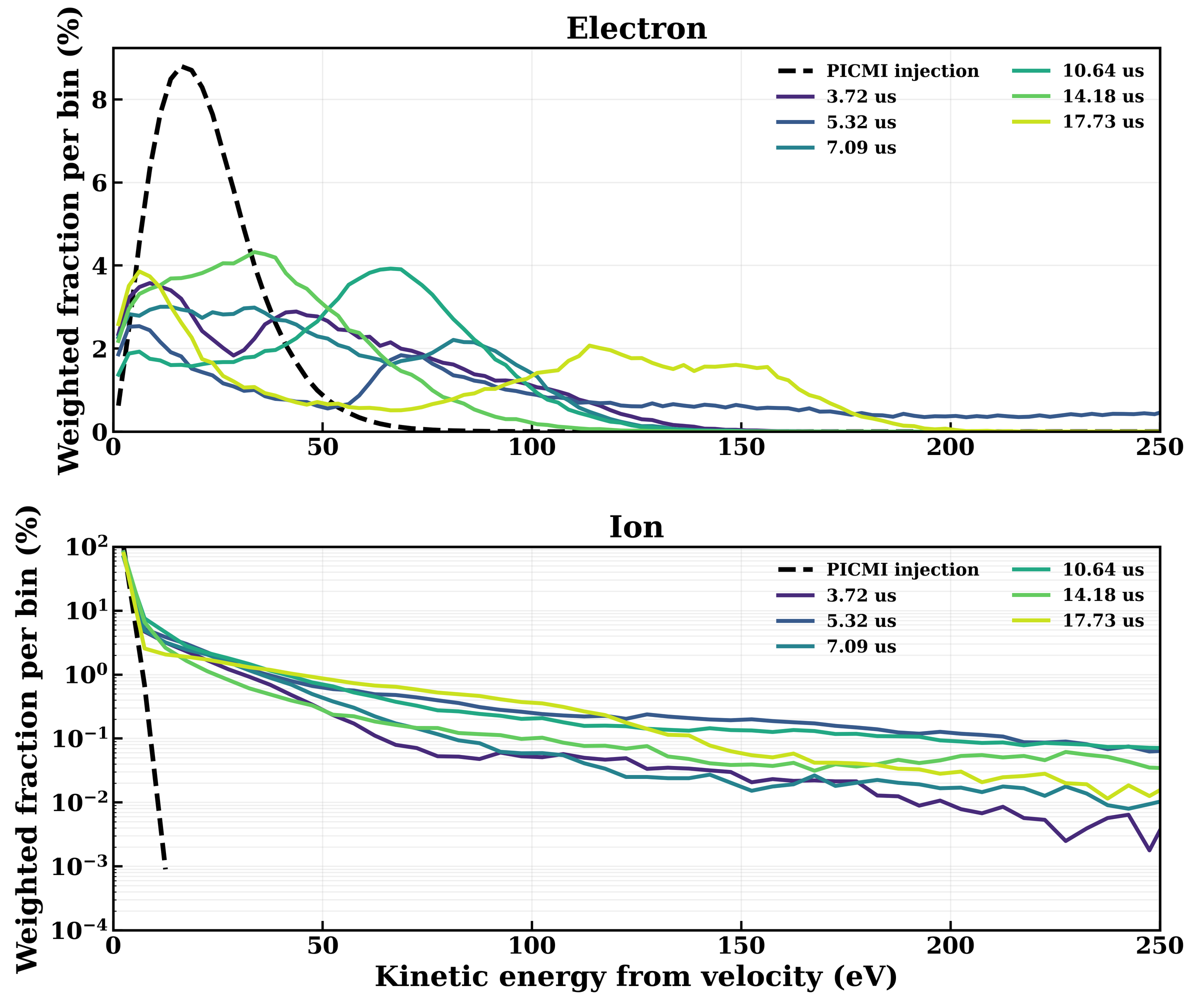}
    \caption{Comparison between the injected ion-energy distribution
    and the self-consistent PIC ion-energy distribution. The injected
    population is confined to low energies, whereas the evolved
    population develops a pronounced energetic tail.}
    \label{fig:injection_selfconsistent_distribution}
\end{figure}

The possible contribution of the imposed inlet population was first
assessed by comparing its initial ion-energy distribution with the
self-consistent PIC population. 
The corresponding energy distributions are shown in
figure~\ref{fig:injection_selfconsistent_distribution}.

The injected ions have a mean kinetic energy of approximately
1.43~eV; 90\% of the injected ions have energies below 2.82~eV and
99\% below 4.71~eV. In contrast, the self-consistent ion population
extends to tens and hundreds of electronvolts. The energetic tail is
therefore absent from the prescribed inlet distribution and develops
during the self-consistent plume evolution.

Particle histories provide a more direct source classification.
Self-consistent ions crossing either $z_{\max}$ or $r_{\max}$ with
boundary-crossing kinetic energies above 50~eV were combined into a
single energetic-outflow subset and classified according to their
particle-source flags, as described in
section~\ref{sec:source_birth_method}. 
Of the $11\,599$ ions in the energetic-outflow subset, 96.93\%
($n=11\,243$) are associated with ionization inside the plume and
3.07\% ($n=356$) with the imposed inlet source. Interior ionization
therefore dominates the energetic-ion population escaping through the
downstream and outer radial boundaries.

\begin{figure*}[t]
    \centering
    \includegraphics[width=1.0\linewidth]
    {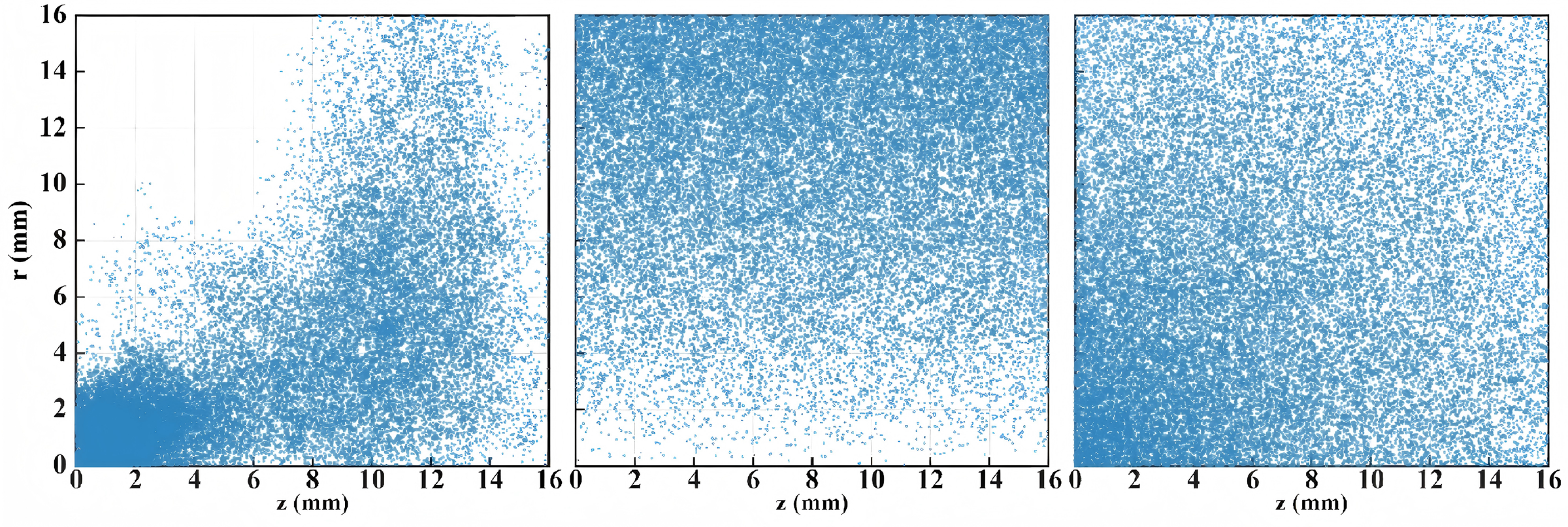}
    \includegraphics[width=1.0\linewidth]
    {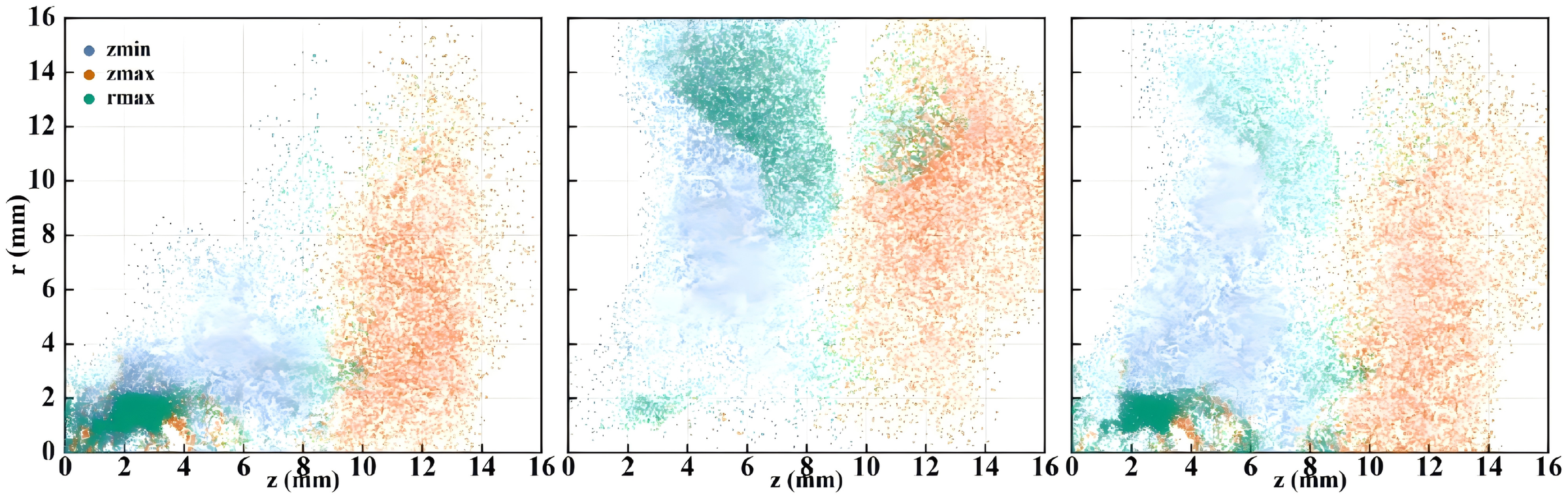}
    \caption{Effect of the prescribed ion birth-position distribution
on energetic-ion formation and escape pathways. The upper row shows
the initial positions of test ions for three source models:
an empirical KDE constructed from PIC ionization events, a distribution
uniform in physical volume, and a neutral-density-weighted distribution
in the $r$--$z$ coordinate plane. The lower row shows the corresponding
birth positions of ions reaching final energies between 50 and 200~eV,
with colors indicating the final escape boundary ($z_{\min}$,
$z_{\max}$, or $r_{\max}$). Each case contains $10^7$ passive ions
propagated through the same PIC-derived time-dependent electric-field
history.}
    \label{fig:birth_model_comparison}
\end{figure*}

The inlet contribution was examined independently by propagating
$5\times10^6$ passive Xe$^+$ ions with the same inlet velocity
distribution as in the self-consistent PIC simulation. 
Of the $5\times10^6$ launched ions, all but 19 returned through
$z_{\min}$; only nine reached $z_{\max}$ and ten reached $r_{\max}$.
Under the sampled field
history, the inlet population therefore has negligible access to
outward downstream or radial trajectories.

The influence of ion birth position was then examined using three source
models: empirical PIC-ionization KDE, uniform physical volume, and
neutral-density-weighted $r$--$z$ sampling. The initial distributions
and the birth positions of ions subsequently reaching 50--200~eV are
shown in figure~\ref{fig:birth_model_comparison}.

Despite their substantially different initial spatial distributions,
all three prescribed birth cohorts contain particles that access
energetic final states. The conditional energetic fractions
$f_{\mathrm{passive}}^{(s,\mathrm{full})}$ are 72.94\%, 43.79\%, and
59.79\% for the empirical-KDE, uniform-volume, and
neutral-density-weighted cohorts, respectively. The corresponding
conditional fractions reaching final energies between 50 and 200~eV
are 56.68\%, 43.43\%, and 54.03\%.

These values quantify conditional energetic-trajectory accessibility
for each prescribed birth cohort rather than the instantaneous
energetic-ion abundance of the self-consistent plasma. Although
energetic trajectories are accessible from all three source models,
their conditional accessibility varies substantially with the
prescribed birth-position distribution.

The energetic subsets also exhibit a consistent spatial organization
with escape boundary: ions reaching $z_{\min}$, $z_{\max}$, and
$r_{\max}$ preferentially originate from different plume regions across
the three prescribed source models. Birth position therefore biases
access to distinct trajectory and escape families rather than uniquely
determining the final particle outcome.

\subsubsection{Electric-field dependence}
\label{sec:field_model_results}

\begin{figure*}[t]
    \centering
    \includegraphics[width=0.98\linewidth]
    {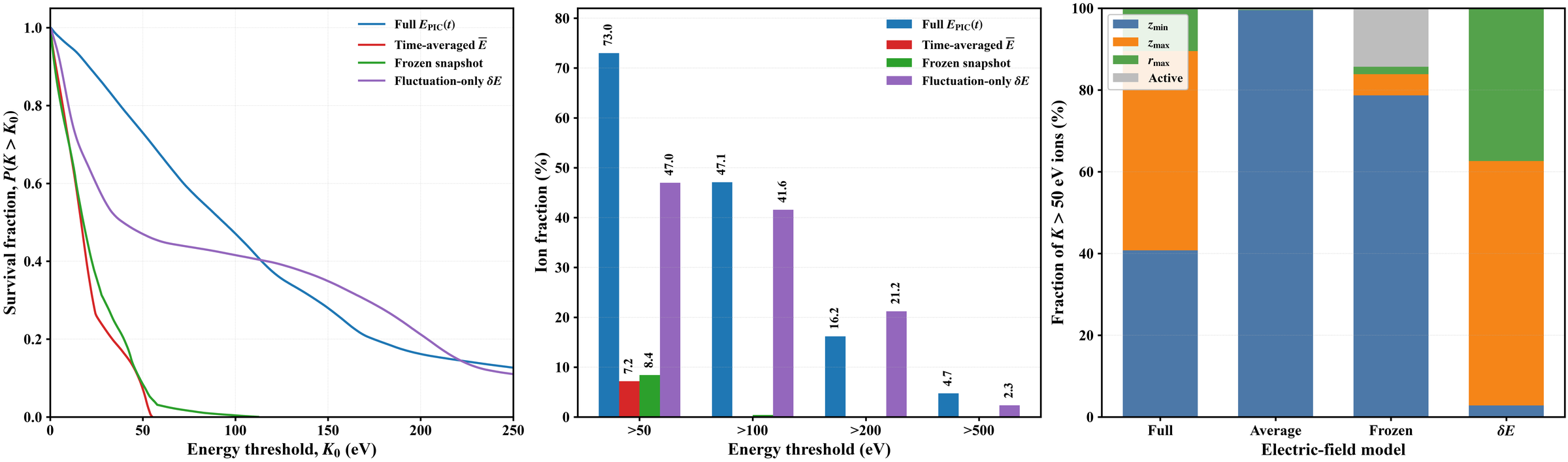}
    \caption{Matched electric-field controls for energetic-ion
    formation using the same empirical-KDE particle realization.
    (a) Survival functions of the final ion energy for the full
    time-dependent PIC field, the time-averaged field, the frozen
    snapshot, and the fluctuation-only field.
    (b) Fractions of ions exceeding selected final-energy thresholds.
    (c) Escape-boundary fractions for ions with $K_f>50$~eV; particles
    remaining inside the computational domain at the end of the
    propagation interval are labeled as active.
    The fluctuation-only field is used as a mechanistic sensitivity
    control and does not represent an independent self-consistent plasma
    solution.}
    \label{fig:field_model_control}
\end{figure*}

The role of the electric-field time history was examined using the
matched empirical-KDE controls described in
section~\ref{sec:field_model_controls}. The same $10^6$ initial
particles were propagated through the full time-dependent PIC field,
the time-averaged field, a frozen PIC snapshot, and the
fluctuation-only field. The resulting energy distributions,
threshold-dependent energetic fractions, and escape statistics are
summarized in figure~\ref{fig:field_model_control}.

Figure~\ref{fig:field_model_control}(a,b) shows that the final-energy
response of the prescribed KDE birth cohort depends strongly on the
electric-field history. For the full time-dependent PIC field, the
conditional energetic fraction defined by
Eq.~\ref{eq:passive_fraction} is
$f_{\mathrm{passive}}^{(\mathrm{KDE},\mathrm{full})}=0.730$.
Thus, 73.0\% of the matched empirical-KDE cohort reaches final energies
above 50~eV, while 47.1\% reaches above 100~eV. For the same initial
particle realization, the time-averaged and frozen fields yield
conditional fractions of only 7.2\% and 8.4\% above 50~eV,
respectively, and essentially no broad population above 100~eV. The
corresponding static fields therefore provide much weaker access to
high-energy trajectories than the full time-dependent field.

The fluctuation-only field also provides substantial access to
high-energy trajectories within the same prescribed cohort. The
corresponding conditional fractions are 47.0\% above 50~eV and 41.6\%
above 100~eV. Its survival function nevertheless differs from that
obtained with the full field. The full PIC field produces a mean final
energy of 135.8~eV, compared with 112.0~eV for the fluctuation-only
case, and retains the larger extreme-energy population.

The conditional passive fractions are not directly comparable with the
instantaneous self-consistent fraction
$f_{\mathrm{SC}}(5.32~\mu\mathrm{s})=0.1187$, because they describe
different statistical ensembles.

Figure~\ref{fig:field_model_control}(c) further shows that the field
model changes the escape pathways as well as the energy distribution.
The time-averaged field directs essentially all energetic ions toward
$z_{\min}$, while the frozen field remains strongly dominated by
$z_{\min}$ escape. In contrast, the full time-dependent field
distributes the energetic population among the axial and radial escape
families, and the fluctuation-only field produces a much larger
contribution from $z_{\max}$ and $r_{\max}$. The temporal field history
therefore affects both energetic-trajectory accessibility and the
eventual escape pathway of the prescribed cohort.

These matched controls show that, for an identical prescribed birth
cohort, the broad final-energy response obtained with the full PIC field
history cannot be reproduced by the corresponding static fields. The
comparison therefore demonstrates strong sensitivity of energetic
trajectory accessibility, final energy, and escape pathway to temporal
field evolution over the analyzed interval. Because changing the
prescribed field also changes the particle trajectory, however, the four
controls are interpreted as mechanistic sensitivity tests rather than as
an additive decomposition of the physical ion-energy gain.

\subsubsection{Trajectory-resolved energy transfer}
\label{sec:electric_work}

\begin{figure*}[t]
    \centering
    \includegraphics[width=0.98\linewidth]
    {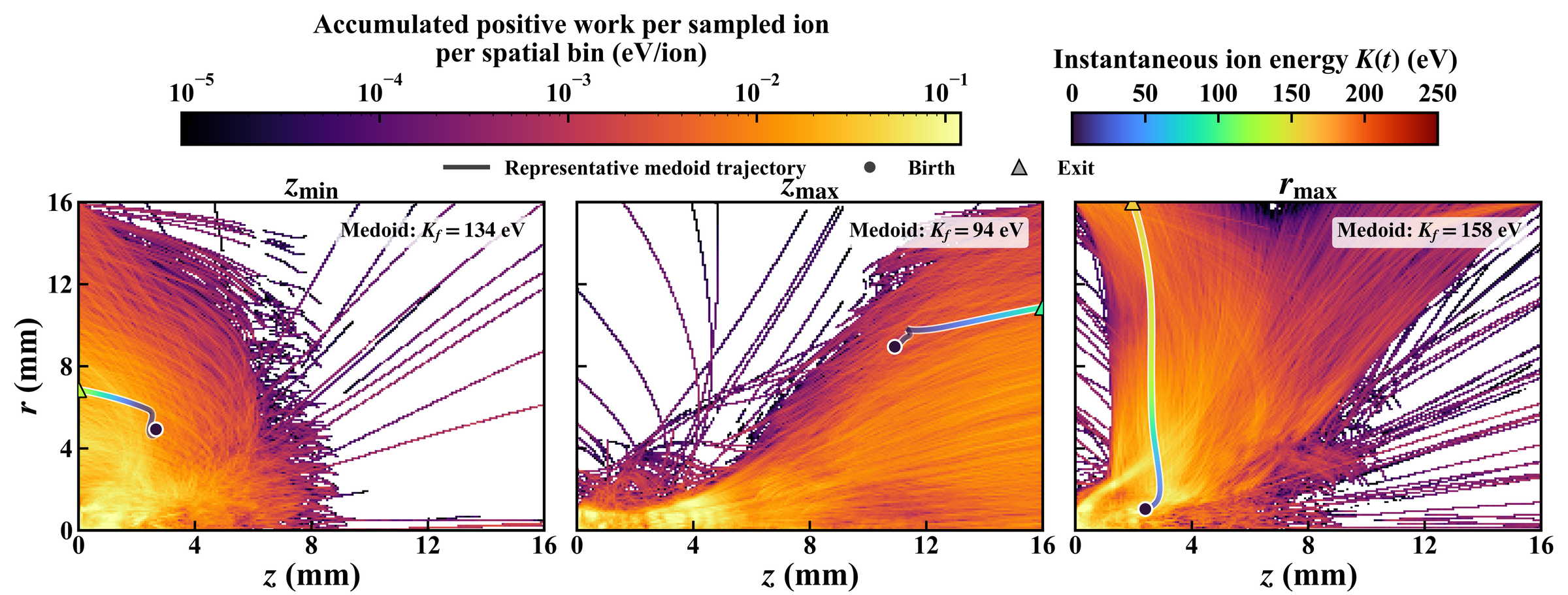}
    \caption{Representative energetic-ion trajectories superimposed on the
spatial distributions of positive electric-field work for ions escaping
through (a) $z_{\min}$, (b) $z_{\max}$, and (c) $r_{\max}$.
The background maps show the accumulated positive work per sampled ion
in each spatial bin, including only trajectory increments satisfying
$q_i\mathbf{E}\cdot\mathbf{v}_p>0$. They therefore identify regions
that preferentially contribute positive kinetic-energy transfer.
For each escape family, a representative medoid trajectory is
superimposed and colored by its instantaneous ion kinetic energy $K_i(t)$.
Filled circles and triangles denote the particle birth and escape
locations, respectively, and the labeled $K_f$ gives the final kinetic
energy of the representative particle.}
    \label{fig:positive_work_hotspot}
\end{figure*}

The particle-level energy transfer for the empirical-KDE cohort from
the birth-position analysis was quantified using the electric-field
work defined in section~\ref{sec:work_method}. To connect the
ensemble-level spatial work distribution with individual particle
dynamics, figure~\ref{fig:positive_work_hotspot} combines positive-work
maps with representative single-particle pathways for the
$z_{\min}$, $z_{\max}$, and $r_{\max}$ escape families. The
representative particles are the pathway medoids selected using the
procedure described in section~\ref{sec:work_method}.

The background distributions show that positive electric-field work is
spatially localized rather than uniformly distributed throughout the
plume. Because these maps contain only positive work increments
accumulated over an ensemble of energetic-ion trajectories, they
identify regions that preferentially contribute positive kinetic-energy
transfer but do not represent either a static acceleration potential or
the net work acquired by an individual ion.

Within $50\leq K_f\leq250$~eV, the dominant work classes are
axial-work dominated for $z_{\min}$ and $z_{\max}$, accounting for
69.99\% and 88.60\% of the corresponding populations, respectively,
and radial-work dominated for $r_{\max}$, accounting for 79.28\%.

The superimposed pathway medoids provide complementary
single-particle views of how ions sample this positive-work landscape.
The representative $z_{\min}$ particle returns toward the upstream
boundary, the $z_{\max}$ particle follows a predominantly downstream
path, and the $r_{\max}$ particle undergoes substantial radial
displacement before leaving the domain. Their final kinetic energies are
134.08, 94.36, and 158.21~eV, respectively. Along each trajectory, the
color represents the instantaneous ion kinetic energy $K_i(t)$, thereby
showing how the particle energy evolves along its actual path from birth
to escape. The three examples illustrate that ions reaching different
escape boundaries traverse different regions of the plume and therefore
sample different histories of the time-dependent electric field.

Consistent with the work-class conditioning used in their selection,
the $z_{\min}$ and $z_{\max}$ pathway medoids have $f_r=0.0712$ and
0.0614, respectively, whereas the $r_{\max}$ medoid has
$f_r=0.9412$. Thus, the selected particles provide illustrative
single-particle realizations of the dominant directional-work classes
for the three escape families. Population-level directional-work trends
are examined independently below.

\begin{figure*}[t]
    \centering
    \includegraphics[width=1.0\linewidth]
    {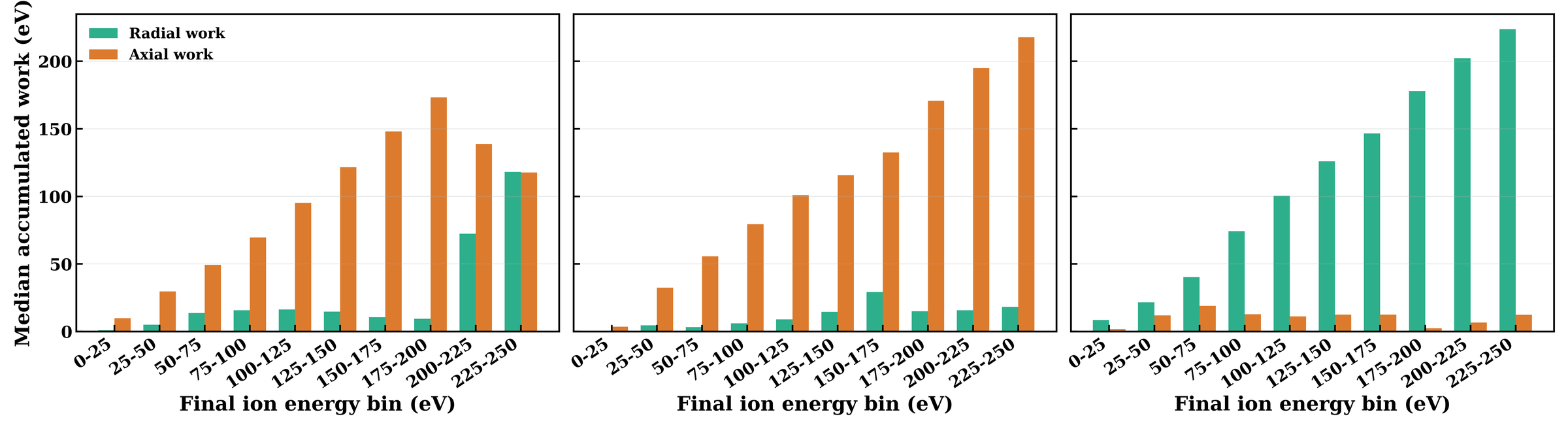}
    \caption{Median radial and axial accumulated electric-field work,
$W_r$ and $W_z$, for passive ions sampled from the empirical
PIC-ionization KDE and propagated through the full time-dependent PIC
field, grouped by final escape boundary and final ion-energy bin.}
    \label{fig:work_components_by_energy}
\end{figure*}

The representative pathways in
figure~\ref{fig:positive_work_hotspot} illustrate how individual ions
can sample different regions and directional components of the
time-dependent electric field, but they do not by themselves establish
population-level behavior. To determine whether these pathway-dependent
differences persist across the energetic-ion ensemble, the accumulated
radial and axial work components were evaluated for the full population
and grouped by final energy and escape boundary, as shown in
figure~\ref{fig:work_components_by_energy}.

The population-level statistics confirm a systematic dependence of the
work partition on escape family. For ions escaping through $z_{\max}$,
the axial contribution $W_z$ provides most of the net positive energy
transfer. For ions reaching $r_{\max}$, the radial contribution $W_r$
becomes increasingly important with final energy and dominates the
higher-energy part of the population. The $z_{\min}$ population shows a less
strongly axial work partition than the $z_{\max}$ population, with an
appreciable radial contribution. These ensemble trends
therefore generalize the pathway differences illustrated by the
representative particles in figure~\ref{fig:positive_work_hotspot}:
different escape families sample different spatial regions and
components of the evolving electric field.

\begin{figure}[t]
    \centering
    \includegraphics[width=1.0\linewidth]
    {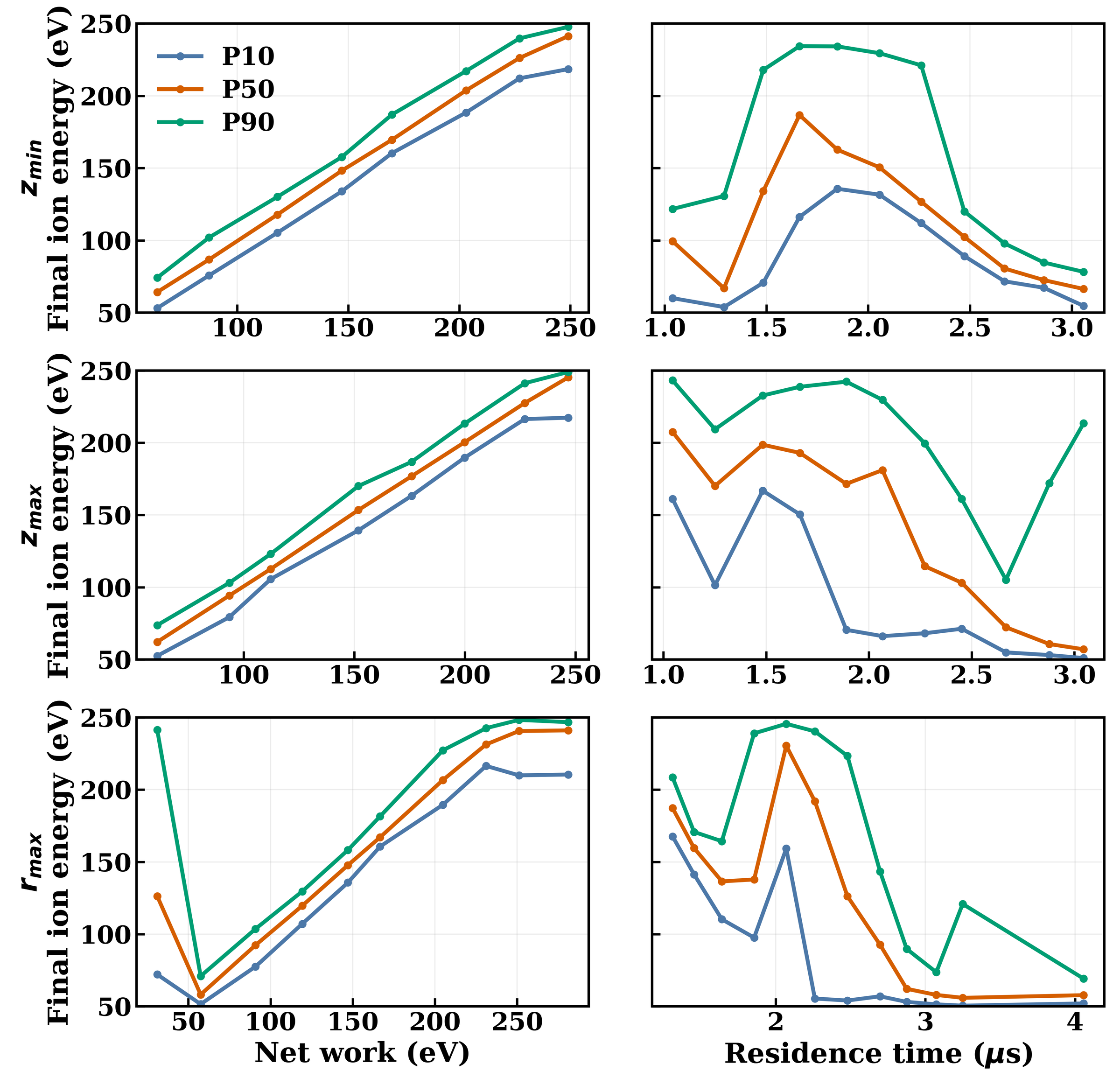}
    \caption{Conditional quantiles of the final ion kinetic energy $K_f$
versus accumulated net electric-field work (left column) and particle
residence time (right column) for passive ions initialized from the
empirical PIC-ionization KDE. Rows correspond to first escape through
$z_{\min}$, $z_{\max}$, and $r_{\max}$. P10, P50, and P90 denote the
10th, 50th, and 90th percentiles of $K_f$ within each bin, whose
horizontal coordinate is the median value of the binned variable.}
    \label{fig:net_work_residence_quantiles}
\end{figure}

Finally, figure~\ref{fig:net_work_residence_quantiles} compares the
final ion energy with accumulated net work and residence time.

The close correspondence between the kinetic-energy change and
$W_{\mathrm{net}}$ is expected from the work--energy relation for the
collisionless passive trajectories and provides a numerical consistency
check on the trajectory integration and field interpolation. 
In contrast, final
energy is not monotonic with residence time. A longer time spent in the
plume is therefore not sufficient to determine whether an ion reaches a
high final energy.

Taken together, the particle-resolved results identify a
source--trajectory--field--work pathway. Interior ionization
dominates the classified energetic outflow, birth position biases
access to distinct trajectory families, temporal electric-field
evolution shapes their subsequent dynamics, and trajectory-integrated
electric-field work accounts for the resulting energy gain. This
particle-level pathway is established without assigning the
time-dependent field to a unique plasma mode.

\section{Discussion}
\label{sec:discussion}

\subsection{Modal interpretation of plume fluctuations}
\label{sec:IAT_discussion}

The present results distinguish two related but separate questions:
whether the fluctuating plume can be assigned to a particular plasma
mode, and how individual ions acquire energy within the resulting
time-dependent electric field. In the experiment, energetic ions
coexist with broadband fluctuations, while the two-probe measurements
do not resolve a continuous ion-acoustic dispersion branch within the
principal apparent-wavenumber interval. Here,
$k_{\mathrm{app}}$ is derived from the principal cross-spectral phase
and is therefore defined modulo $2\pi/\Delta x$. Spectral weight near
$k_{\mathrm{app}}\simeq0$ consequently indicates a small inter-probe
phase difference modulo $2\pi$, rather than uniquely identifying a
true long-wavelength mode with $k_{\mathrm{true}}\simeq0$.
Shorter-wavelength fluctuations may be folded into the same interval,
while spatially coherent or superposed components may also contribute.
The corresponding PIC spectrum is likewise broadband, but is used here
to characterize the collective dynamics of the present kinetic plume
rather than to establish the absence of a particular mode. The
available measurements and simulation therefore do not provide an
unambiguous modal attribution for the fluctuating field.

This distinction is also important when comparing the present
measurements with hollow-cathode experiments in which a clearer
ion-acoustic dispersion branch has been reported. For example,
Miao \emph{et al.} used a 2~mm probe separation and maintained a
keeper current of $I_k=1.5$~A during their measurements, whereas the
present experiment uses a 3~mm probe separation and the keeper is
electrically floating after ignition
\cite{miao2025experimental}. The different probe separations directly
change the accessible principal wavenumber interval, while the keeper
condition constitutes a different electrical boundary for plume
current closure and local potential adjustment. Previous work has also
shown that keeper--plasma interaction can modify hollow-cathode plume
oscillations \cite{meng2019triggering}. These differences may therefore
contribute to different observed fluctuation regimes, although the
present measurements do not isolate the keeper boundary condition as a
causal control parameter.

A related PIC configuration was previously interpreted in terms of a
charge-separation instability \cite{zhao2025generation}. That study
addressed the collective origin and evolution of the large-amplitude
self-consistent potential structures in its numerical configuration.
The present PIC case, however, uses modified domain dimensions,
boundary potentials, injection parameters, and neutral-density profile.
The collective-mode attribution established for the earlier
configuration is therefore not assumed \emph{a priori} for the present
case. A dedicated collective-mode analysis would be required to
determine whether the same interpretation applies to the modified
configuration. The particle-resolved analysis developed here instead
asks a complementary question: given the resulting time-dependent
self-consistent field, how do ion source, trajectory accessibility, field
history, and electric-field work combine to produce energetic particle
pathways?

This distinction is important because the dynamical role of a
time-dependent electric field does not depend on first identifying its
modal origin. For an identical prescribed particle cohort, the matched
field controls show that replacing the full time-dependent PIC field by
its time-averaged or frozen counterpart strongly suppresses access to
high-energy trajectories. The fluctuation-only field retains substantial
energetic-trajectory accessibility, although its particle response
differs from that obtained with the full field. 
These comparisons demonstrate that, over the analyzed interval,
temporal field evolution strongly influences particle dynamics, but do
not determine whether the underlying fluctuations arise
from ionization dynamics, ion-acoustic activity, other instabilities, or
their nonlinear interaction.
Ion-acoustic activity and larger-scale time-dependent electrostatic
dynamics therefore need not be regarded as mutually exclusive
contributors to the time-dependent plume field. They may coexist with
different relative importance, whereas the energy acquired by an
individual ion is determined by the electric field sampled along its
actual trajectory.

The trajectory-resolved work analysis addresses a different level of
the problem. The quantity
$q_i\mathbf{E}\cdot\mathbf{v}_p$
gives the instantaneous electric-field power transferred to an ion, and
its trajectory integral accounts for the corresponding kinetic-energy
change. This provides a particle-level description of where and through
which field components energy is transferred, but does not identify the
plasma mode responsible for generating the field. Modal attribution and
particle-level energy transfer should therefore be treated as distinct
diagnostic questions.

The principal conclusion is consequently not that ion-acoustic
activity is absent, nor that the energetic-ion population can be
uniquely assigned to an alternative collective mode.
Shorter-wavelength, intermittent, off-axis, or spatially localized
ion-acoustic activity may remain unresolved by the present diagnostics,
and previous studies have established well-developed IAT under other
hollow-cathode operating conditions. Rather, the present results show
that a consistent particle-level energetic-ion pathway can be
identified without first assigning the time-dependent field to a
unique resolved plasma mode. This interpretation is compatible with a
regime-dependent picture in which ion-acoustic activity and
larger-scale electrostatic dynamics may coexist with different relative
importance under different operating and boundary conditions.

% ============================================================
\subsection{Source--trajectory--field--work pathway}
\label{sec:implications}

The particle-resolved results indicate that energetic-ion formation is
a history-dependent kinetic process involving coupled source,
trajectory, and field effects. Final ion-energy distributions alone do
not retain this history: ions with similar final energies may originate
from different regions, follow different escape pathways, and sample
different radial and axial electric-field contributions. The combined
analyses therefore organize energetic-ion formation into a
\emph{source--trajectory--field--work pathway}.

The source and birth-position analyses define the first two elements of
this pathway. Interior ionization dominates the classified energetic
outflow, while the inlet control shows that the imposed inlet population
has little access to the outward trajectories sampled here. However,
energetic trajectories remain accessible from all three prescribed
birth-position distributions. Source localization therefore does not
uniquely determine the final particle state; instead, it biases the
statistical accessibility of different trajectory and escape families.

The matched controls then separate this source dependence from the
effect of the subsequent field history. In the birth-position
calculations, the prescribed field is held fixed while the spatial
source distribution is varied. In the field controls, the initial
particle realization is held fixed while the electric-field history is
changed. Together, these comparisons show that energetic-trajectory
accessibility depends both on the particle's initial location and on the
subsequent temporal evolution of the field.

These two influences are nevertheless dynamically coupled rather than
strictly sequential. Changing the prescribed field changes the particle
trajectory, which in turn changes the sequence of field regions sampled
along the path. The full, mean, frozen, and fluctuation-only fields
therefore do not represent independent or additive acceleration
channels. Instead, the particle trajectory and the time-dependent field
jointly determine the history of particle--field interaction.

Trajectory-resolved work provides the corresponding energy accounting.
The positive-work maps show that energy transfer is concentrated in
localized regions of the plume, while the representative pathway
medoids illustrate how individual ions traverse these regions between
birth and escape and how their kinetic energy $K_i(t)$ evolves along
the resulting path. These representative trajectories are illustrative
single-particle realizations selected from the dominant directional-work class of each escape family and are not used by themselves to
infer population-level directional-work behavior.

The systematic directional dependence is instead established by the
ensemble statistics of $W_r$ and $W_z$. These statistics show that the
relative radial and axial work contributions vary among escape
families, consistent with the distinct pathway geometries illustrated
by the representative particles. Residence time alone does not show a
corresponding monotonic relation with final energy.

Within this framework, source localization biases the initial
statistical occupation of the plume, birth position influences access
to different trajectory families, temporal electric-field evolution
shapes the subsequent particle dynamics, and trajectory-integrated
electric-field work accounts for the resulting kinetic-energy change.
The framework connects ion production, transport, and particle-level
energy transfer without requiring either a single stationary
acceleration structure or a unique modal attribution of the fluctuating
field.

This interpretation may also be relevant to hollow-cathode operation
and lifetime. Energetic-particle fluxes incident on the keeper,
downstream components, or surrounding structures may depend not only on
the global plasma-potential distribution but also on the spatial
ionization pattern and on how the evolving plume field connects source
regions to axial and radial escape pathways. Operating conditions that
modify either source localization or temporal field evolution may
therefore alter both the energetic-ion population and its spatial
distribution.

% ============================================================
\subsection{Limitations}
\label{sec:limitations}

Several limitations define the scope of the present conclusions. First,
the PIC calculation is used as a representative kinetic reference
rather than as a quantitative reconstruction of a specific experimental
operating point. The experiment and simulation also sample different
physical quantities and particle populations. The RPA measures an
acceptance-limited ion population, whereas the numerical energy
distributions are sampled at selected computational boundaries, and the
experimental potential fluctuation is inferred from a probe-derived
proxy whereas the numerical potential is obtained directly from the
electrostatic PIC solution. The experiment--simulation comparisons are
therefore interpreted qualitatively rather than as point-by-point
validation.

Second, the experimental modal interpretation is limited by the
two-probe diagnostic geometry. The finite probe separation restricts
the principal apparent-wavenumber interval, and the inferred
$k_{\mathrm{app}}$ is obtained from a cross-spectral phase defined
modulo $2\pi$. Consequently, spectral weight near
$k_{\mathrm{app}}=0$ cannot uniquely distinguish a true
long-wavelength fluctuation from a shorter-wavelength component folded
into the principal interval. The acoustic-speed curve additionally
provides only a cold-ion, zero-drift reference. The measurements
therefore do not exclude shorter-wavelength, intermittent, spatially
localized, off-axis, or Doppler-shifted ion-acoustic activity that is
not uniquely resolved within the present configuration. Multiple probe
separations or an array-based spatial measurement would be required for
more robust phase unwrapping and modal identification.

The passive-particle calculations introduce additional limitations.
They are one-way coupled and collisionless: the particles respond to
the stored PIC electric field but do not modify the self-consistent
plasma state or undergo collisional interactions, including
Xe$^+$--Xe charge exchange. These calculations are therefore
interpreted as controlled diagnostics of trajectory accessibility,
field-history sensitivity, and electric-field energy transfer rather
than as an independent reconstruction of the self-consistent ion
population. In addition, the present analysis uses the selected
$3.7259$--$8.7259~\mu\mathrm{s}$ field interval. Broader temporal
sampling would be required to quantify the variability of the inferred
trajectory and field-history responses over the full plume evolution.

Finally, the mean, frozen, and fluctuation-only fields are mechanistic
controls rather than independent self-consistent plasma states, and
their particle-energy responses are not additive contributions to the
full-field result. The underlying PIC model is two-dimensional,
axisymmetric, and electrostatic, with a prescribed neutral background;
three-dimensional structure, electromagnetic effects, and neutral
evolution may modify the ionization distribution, electric-field
dynamics, and accessible trajectories. The trajectory-resolved work
analysis identifies how energy is transferred to individual ions but
does not determine the modal origin of the responsible field. Resolving
that origin would require additional spatially resolved fluctuation
measurements and corresponding kinetic analysis.

\section{Conclusions}
\label{sec:conclusions}

Energetic-ion formation in a low-current hollow-cathode plume was
investigated using experiments, self-consistent electrostatic
particle-in-cell simulation, and particle-resolved analysis. The
experiment shows that a substantial energetic-ion population coexists
with broadband and structured plume fluctuations over discharge
currents of $0.8$--$3.5$~A. Two-point phase-derived
frequency--wavenumber measurements do not resolve a continuous
ion-acoustic dispersion branch within the principal
apparent-wavenumber interval. Because the inferred
$k_{\mathrm{app}}$ is obtained from a wrapped cross-spectral phase,
this result does not establish the absence of ion-acoustic activity;
rather, it shows that the available fluctuation diagnostics do not
provide an unambiguous modal attribution for the energetic-ion
population. The modified representative PIC case likewise develops
broadband time-dependent electrostatic fluctuations together with a
nonthermal energetic-ion population and provides the particle and field
histories required for kinetic analysis.

The particle-resolved analysis identifies a
source--trajectory--field--work pathway for energetic-ion formation.
Source classification shows that energetic ions escaping through the
downstream and outer radial boundaries are predominantly generated by
ionization inside the plume, while controlled birth-position
calculations show that source localization biases access to distinct
trajectory and escape families. Matched electric-field controls further
show that, for an identical prescribed birth cohort, replacing the full
time-dependent PIC field by its time-averaged or frozen counterpart
strongly suppresses access to high-energy trajectories over the
analyzed interval. The temporal evolution of the electric field
therefore affects not only the final ion-energy distribution but also
the set of trajectories and escape pathways accessible to the
particles.

At the single-particle level, the direct kinetic-energy gain is supplied
by work done by the self-consistent time-dependent electrostatic field
along the actual ion trajectory, as quantified by
$\Delta K_i=\int q_i\mathbf{E}\cdot\mathbf{v}_i\,dt$. The radial and
axial contributions to this work differ systematically among the
different escape families. Ions escaping through $z_{\max}$ acquire
most of their net positive energy through axial electric-field work,
whereas radial work becomes increasingly important for ions escaping
through $r_{\max}$. In contrast, residence time alone does not determine
the final ion energy. Energetic-ion formation is therefore not
described by residence time or by a single stationary acceleration
structure alone; it depends on whether an ion accesses trajectories
that sample favorable regions and phases of the evolving electrostatic
field.

These results provide a particle-level picture in which source
localization biases the initial statistical occupation of the plume,
birth position influences trajectory accessibility, temporal
electric-field evolution shapes the subsequent particle dynamics, and
trajectory-integrated electrostatic-field work directly accounts for
the resulting kinetic-energy change. The particle-level energization
pathway can therefore be identified without first assigning the
time-dependent plume field to a unique resolved plasma mode. This
conclusion neither excludes unresolved ion-acoustic activity nor
requires ion-acoustic and larger-scale electrostatic dynamics to be
mutually exclusive. Instead, the results are compatible with a
regime-dependent picture in which different collective processes may
contribute to generating the time-dependent plume field, while the
direct particle-level energy transfer is determined by the
trajectory-integrated electrostatic-field work.

\ack{The authors acknowledge financial support from the National Natural Science Foundation of China (Grant Nos. U22B20120 and 52472403).}
 
% \funding{Sample text inserted for demonstration.}
% % This section is a list of funder names and grant numbers

% \roles{Sample text inserted for demonstration.}
% List author names and the contributions made to the article, using terms from the NISO Contributor Roles Taxonomy (CRediT) https://credit.niso.org

% \data{Sample text inserted for demonstration.}
% % For more information on IOP Publishing's research data policy see: https://publishingsupport.iopscience.iop.org/questions/research-data/

 \data{The data that support the findings of this study are available from the corresponding author upon reasonable request.}

% \section*{References}
\bibliographystyle{iopart-num}
\bibliography{reference}

@article{zhao2025generation,
  title={Generation of anomalously energetic ions in hollow cathode plume via charge separation instability},
  author={Zhao, Yinjian and Wang, Baisheng and Meng, Tianhang},
  journal={Plasma Science and Technology},
  volume={27},
  number={12},
  pages={125503},
  year={2025}
}

@inproceedings{fedeli2022pushing,
  title={Pushing the frontier in the design of laser-based electron accelerators with groundbreaking mesh-refined particle-in-cell simulations on exascale-class supercomputers},
  author={Fedeli, Luca and Huebl, Axel and Boillod-Cerneux, France and Clark, Thomas and Gott, Kevin and Hillairet, Conrad and Jaure, Stephan and Leblanc, Adrien and Lehe, R{\'e}mi and Myers, Andrew and others},
  booktitle={SC22: international conference for high performance computing, networking, storage and analysis},
  pages={1--12},
  year={2022},
  organization={IEEE}
}

@article{jorns2014ion,
  title={Ion acoustic turbulence in a 100-A LaB 6 hollow cathode},
  author={Jorns, Benjamin A and Mikellides, Ioannis G and Goebel, Dan M},
  journal={Physical Review E},
  volume={90},
  number={6},
  pages={063106},
  year={2014},
  publisher={APS}
}

@inproceedings{chen2003mini,
  title={Mini-course on plasma diagnostics},
  author={Chen, Francis F},
  booktitle={IEEE-ICOPS Meeting, Jeju, Korea},
  volume={5},
  year={2003}
}

@article{miao2025experimental,
  title={Experimental insights into discharge instability and energetic ion dynamics in ampere-level hollow cathodes},
  author={Miao, Long and Jia, Jintao and Tian, Feng and Gu, Zengjie and Hou, Xiao},
  journal={Plasma Sources Science and Technology},
  volume={34},
  number={10},
  pages={105008},
  year={2025},
  publisher={IOP Publishing}
}

@article{goebel2007potential,
  title={Potential fluctuations and energetic ion production in hollow cathode discharges},
  author={Goebel, Dan M and Jameson, Kristina K and Katz, Ira and Mikellides, Ioannis G},
  journal={Physics of Plasmas},
  volume={14},
  number={10},
  year={2007},
  publisher={AIP Publishing}
}

@article{goebel2021plasma,
  title={Plasma hollow cathodes},
  author={Goebel, Dan M and Becatti, Giulia and Mikellides, Ioannis G and Lopez Ortega, Alejandro},
  journal={Journal of Applied Physics},
  volume={130},
  number={5},
  year={2021},
  publisher={AIP Publishing}
}

@article{patterson1999generation,
  title={The generation of high energy ions in hollow cathode discharges},
  author={Patterson, Stephen W and Fearn, David G},
  journal={IEPC Paper},
  pages={99--125},
  year={1999}
}

@article{lev2019recent,
  title={Recent progress in research and development of hollow cathodes for electric propulsion},
  author={Lev, Dan R and Mikellides, Ioannis G and Pedrini, Daniela and Goebel, Dan M and Jorns, Benjamin A and McDonald, Michael S},
  journal={Reviews of Modern Plasma Physics},
  volume={3},
  number={1},
  pages={6},
  year={2019},
  publisher={Springer}
}

@article{goebel2005energetic,
  title={Energetic ion production and keeper erosion in hollow cathode discharges},
  author={Goebel, Dan M and Jameson, Kristina and Katz, Ira and Mikellides, Ioannis G and Polk, J},
  journal={IEPC Paper},
  volume={266},
  pages={2005},
  year={2005}
}

@article{zhao2026review,
  title={A review of discharge instabilities in hollow cathodes},
  author={Zhao, Yinjian and Wang, Baisheng and Meng, Tianhang and Ning, Zhongxi and Yu, Daren},
  journal={Plasma Science and Technology},
  year={2026}
}

@phdthesis{georgin2020ionization,
  title={Ionization instability of the hollow cathode plume},
  author={Georgin, Marcel},
  school = {University of Michigan},
  year={2020}
}

@article{meng2019triggering,
  title={Triggering of ionization oscillations in hollow cathode discharge by keeper electrode},
  author={Meng, Tianhang and Ning, Zhongxi and Yu, Daren},
  journal={Physics of Plasmas},
  volume={26},
  number={9},
  year={2019},
  publisher={AIP Publishing}
}

@inproceedings{jorns2016first,
  title={First-principles modelling of the IAT-driven anomalous resistivity in hollow cathode discharges I: Theory},
  author={Jorns, Benjamin and Lopez Ortega, Alejandro and Mikellides, Ioannis G},
  booktitle={52nd AIAA/SAE/ASEE Joint Propulsion Conference},
  pages={4626},
  year={2016}
}

@article{georgin2019correlation,
  title={Correlation of ion acoustic turbulence with self-organization in a low-temperature plasma},
  author={Georgin, Marcel P and Jorns, Benjamin A and Gallimore, Alec D},
  journal={Physics of Plasmas},
  volume={26},
  number={8},
  year={2019},
  publisher={AIP Publishing}
}

@article{imai2022effect,
  title={The Effect of Discharge Mode on Ion Energy and Plasma Potential in the Plume Plasma Region},
  author={Imai, Shun and Imaguchi, Daisuke and Watanabe, Hiroki and Kubota, Kenichi and Cho, Shinatora and Oshio, Yuya and Funaki, Ikkoh},
  journal={TRANSACTIONS OF THE JAPAN SOCIETY FOR AERONAUTICAL AND SPACE SCIENCES},
  volume={65},
  number={1},
  pages={1--10},
  year={2022},
  publisher={THE JAPAN SOCIETY FOR AERONAUTICAL AND SPACE SCIENCES}
}

@article{wang2022presence,
  title={Presence of energetic ions in hollow cathode discharge with low frequency oscillations},
  author={Wang, Fu-Feng and Meng, Tian-Hang and Yu, Da-Ren and Ning, Zhong-Xi and Zhu, Xi-Ming},
  journal={Journal of Physics D: Applied Physics},
  volume={55},
  number={45},
  pages={455202},
  year={2022},
  publisher={IOP Publishing}
}

@article{luo2024coupled,
  title={A coupled plasma-thermal model for hollow cathode power decomposition and parametric analysis},
  author={Luo, Zihao and Ren, Junxue and Cao, Lehui and Zhang, Guangchuan and Wang, Yibai and Zhang, Zun and Wang, Weizong and Tang, Haibin},
  journal={Plasma Sources Science and Technology},
  volume={33},
  number={10},
  pages={105020},
  year={2024},
  publisher={IOP Publishing}
}

@article{beall1982estimation,
  title={Estimation of wavenumber and frequency spectra using fixed probe pairs},
  author={Beall, JM and Kim, YC and Powers, EJ},
  journal={Journal of Applied Physics},
  volume={53},
  number={6},
  pages={3933--3940},
  year={1982},
  publisher={American Institute of Physics}
}

@inproceedings{katz2006production,
  title={Production of high energy ions near an ion thruster discharge hollow cathode},
  author={Katz, Ira and Mikellides, Ioannis and Goebel, Dan and Jameson, Kristina and Wirz, Richard and Johnson, Lee},
  booktitle={42nd AIAA/ASME/SAE/ASEE Joint Propulsion Conference \& Exhibit},
  pages={4485},
  year={2006}
}

@article{williams2000laser,
  title={Laser-induced fluorescence characterization of ions emitted from hollow cathodes},
  author={Williams, George J and Smith, Timothy B and Domonkos, Matthew T and Gallimore, Alec D and Drake, R Paul},
  journal={IEEE Transactions on Plasma Science},
  volume={28},
  number={5},
  pages={1664--1675},
  year={2000},
  publisher={IEEE}
}

@article{foster2005downstream,
  title={Downstream ion energy distributions in a hollow cathode ring cusp discharge},
  author={Foster, John E and Patterson, Michael J},
  journal={Journal of propulsion and power},
  volume={21},
  number={1},
  pages={144--151},
  year={2005}
}

@inproceedings{jorns2014investigation,
  title={Investigation of energetic ions in a 100-a hollow cathode},
  author={Jorns, Benjamin and Goebel, Dan M and Mikellides, Ioannis G},
  booktitle={50th AIAA/ASME/SAE/ASEE Joint Propulsion Conference},
  pages={3826},
  year={2014}
}

@article{dodson2019measurements,
  title={Measurements of ion velocity and wave propagation in a hollow cathode plume},
  author={Dodson, Christopher and Jorns, Benjamin and Wirz, Richard},
  journal={Plasma Sources Science and Technology},
  volume={28},
  number={6},
  pages={065009},
  year={2019},
  publisher={IOP Publishing}
}

\end{document}